*To be submitted to Phys. Rev.* **B**

# Piezoresponse Force Spectroscopy of Ferroelectric Materials

Anna N. Morozovska and Sergei V. Svechnikov

V. Lashkaryov Institute of Semiconductor Physics, National Academy of Science of Ukraine,

41, pr. Nauki, 03028 Kiev, Ukraine

Eugene A. Eliseev

Institute for Problems of Materials Science, National Academy of Science of Ukraine,

3, Krjijanovskogo, 03142 Kiev, Ukraine

Stephen Jesse, Brian J. Rodriguez, and Sergei V. Kalinin[*,†]

Materials Sciences and Technology Division, Oak Ridge National Laboratory,

Oak Ridge, TN 37831

## Abstract

Piezoresponse Force Spectroscopy (PFS) has emerged as a powerful technique for probing highly localized switching behavior and the role of microstructure and defects on switching. The application of a dc bias to a scanning probe microscope tip in contact with a ferroelectric surface results in the nucleation and growth of a ferroelectric domain below the tip, resulting in changes in local electromechanical response. Resulting hysteresis loops contains information on local ferroelectric switching behavior. The signal in PFS is the convolution of the volume of the nascent domain and the probing volume of the tip. Here, we analyze the signal formation mechanism in PFS by deriving the main parameters of domain nucleation in a semi-infinite material and establishing the relationships between domain parameters and PFM signal using a linear Green's function theory. The effect of surface screening and finite Debye length on the switching behavior is established. In particular, we predict that the critical nucleus size in PFM is controlled by the surface screening mechanism and in the absence of screening, tip-induced switching is impossible. Future prospects of PFS to study domain nucleation in the vicinity of defects, local switching centers in ferroelectrics, and unusual polarization states in low-dimensional ferroelectrics are discussed.

[*] Corresponding author, sergei2@ornl.gov

[†] Also with the Center of Nanophase Materials Sciences, ORNL, and Department of Materials Science and Engineering, North Carolina State University, Raleigh, NC

## I. Introduction

In the last decade, ferroelectric materials have attracted much attention for electronic device applications such as non-volatile memories[1], ferroelectric data storage devices,[2, 3] or as a platform for nanofabrication.[4] This has stimulated a number of theoretical and experimental studies of ferroelectric properties in low dimensional systems,[5,6] including the size limit for ferroelectricity[7, 8, 9] and intrinsic switching[10] in thin films, and unusual polarization ordering in ferroelectric nanoparticles and nanowires.[11, 12] Further progress in these fields necessitate fundamental studies of ferroelectric domain structures and polarization switching phenomena on the nanoscale. Strong coupling between the local polarization and piezoelectric response in ferroelectric materials allows using the latter as the physical basis for domain structure imaging and probing ferroelectric phenomena down to the nanometer scale level. In the last decade, the invention of Piezoresponse Force Microscopy (PFM)[13,1415,16] has enabled sub-10 nanometer resolution imaging of crystallographic and molecular orientations, surface termination, and domain structures in ferroelectric and piezoelectric materials. In materials with switchable polarization, the smallest domain size reported to date is 5 nm (see Ref. 17) and local polarization patterning down to 8 nm has been demonstrated,[19] within an order of magnitude of the atomic limit. Finally, local electromechanical hysteresis loop measurements (Piezoresponse force spectroscopy) have been developed,[16] providing insight into local switching behavior and mechanisms of polarization switching on the nanoscale.

The characteristic shapes of the electromechanical hysteresis loops in PFM and macroscopic polarization-electric (*P*-*E*) field loops are similar, resulting in a number of attempts to interpret PFM hysteresis loops in terms of macroscopic materials properties. However, in the macroscopic case, the loop shape is determined by the collective processes in ferroelectric ceramics, single crystals, or thin films, including reversible and irreversible displacements of existing domain walls,[18] wall interactions with grain boundaries, defects, and strain fields, nucleation of new domains, and domain growth.[19, 20] Depending on the exact nature of the system (ceramics vs. thin film), relevant materials parameters, including Preisach distributions of the switching fields, nucleation and domain growth rates etc. can be extracted from the hysteresis loop shape. However, in all cases, the hysteresis loop shape provides information on the switching behavior in a uniform or nearly uniform electric field, averaged over macroscopic volumes and comprising multiple switching events, thus providing statistical characteristics of the switching process.

Conversely, in the PFM experiment, the electric field is highly localized in the vicinity of the atomic force microscope (AFM) tip, with the maximum value achieved at the tip-surface contact. Therefore, domain nucleation is initiated directly below the tip, with subsequent vertical and lateral domain growth. This scenario has been supported by numerous experimental and theoretical studies of local ferroelectric domain switching.[21, 22, 23, 24, 25] The interpretation of the PFM spectroscopic data is complicated by the fact that the signal generation volume is also localized in the vicinity of the tip and is determined by the geometric characteristics of the tip-surface system. The PFM hysteresis loop shape is thus determined by the convolution of the signal generation volume and the shape of nascent domain. Hence, despite the qualitative similarity between the hysteresis loop shape in macroscopic and microscopic cases, the fundamental mechanisms behind the loop formation are fundamentally different, necessitating the quantitative analysis of local electromechanical hysteresis loop formation in PFM.

Here, we develop the theoretical background for PFM hysteresis measurements. General principles of PFM and the existing results and models for the interpretation of Piezoresponse Force Spectroscopy are summarized in Section II. The thermodynamics of domain switching and role of surface screening and finite material conductivity is analyzed in Section III. The relationship between geometric parameters of a domain and the PFM signal is derived Section IV. The experimental results are briefly discussed in Section V, and the role of pinning on hysteresis loop formation is discussed. We also demonstrate that PFM spectroscopy can provide information on the local mechanism for domain nucleation and the thermodynamic parameters of the switching process, and discuss the future potential of PFM spectroscopy to probe nanoscale ferroelectric phenomena.

## II. Review of current results on PFM switching studies

### II.1. Phenomenological studies of PFS

Piezoresponse Force Microscopy is based on the detection of bias-induced piezoelectric surface deformation. The tip is brought into contact with the surface and the piezoelectric response of the surface is detected as the first harmonic component of bias-induced tip deflection, $u = u_0 + A\cos(\omega t + \varphi)$. The phase of the electromechanical response of the surface, $\varphi$, yields information on the polarization direction below the tip. For $c^-$ domains (polarization vector pointing downward) the application of a positive tip bias results in the expansion of the sample and surface oscillations are in phase with the tip voltage, φ = 0. For $c^+$ domains, φ = 180°. The piezoresponse amplitude, $PR = A/V_{ac}$, defines the local electromechanical activity of the surface.

One of the key questions in understanding ferroelectric materials are the mechanisms for polarization switching and the role of defects, vacancies, domain walls, and other microstructural elements on switching processes. A closely related issue is the nature of the defect sites that allow domain nucleation at low electric fields (Landauer paradox).[26] The application of dc bias to the PFM tip can result in local polarization switching below the tip, thus enabling the creation of domains, which can subsequently be imaged in real space. Studies of domain evolution with time or bias provide insight into the switching behavior and the effect of defects and disorder on switching. Recent studies by Gruverman *et al.* have shown that domain nucleation in ferroelectric capacitors during repetitive switching cycles is always initiated at the same defect regions;[27] similarly, the grain boundaries were shown to play an important role in domain wall pinning.[28] Paruch *et al.* have used local studies of domain growth kinetics[29] and domain wall morphology[30] to establish the origins of disorder in ferroelectric materials. Dawber *et al* interpreted the non-uniform wall morphologies as evidence for skyrmion emission during domain wall motion.[31] Most recently, Agronin *et al.* have observed domain pinning on structural defects.[32]

The primary limitation of these studies of domain growth is the large time required to perform multiple switching and imaging steps. Moreover, the information is obtained on the domain growth initiated at a single point for different bias conditions, thus precluding systematic studies of microstructure influence on domain growth process. An alternative approach to study domain dynamics in the PFM experiment is based on local spectroscopic measurements, in which the domain switching and electromechanical detection are performed simultaneously, yielding a local electromechanical hysteresis loop. In-field hysteresis loop measurements were first reported by Birk *et al.* [33] using an STM tip and Hidaka *et al.*[16] using an AFM tip. In this method, the response is measured simultaneously with the application of the dc electric field, resulting in an electrostatic contribution to the signal. To avoid this problem, a technique to measure remanent loops was reported by Guo *et al*.[34] In this case, the response is determined after the dc bias is removed, minimizing the electrostatic contribution to the signal. However, after the bias is turned off, domain relaxation is possible.

In a parallel development, Roelofs *et al*.[35] demonstrated the acquisition of both vertical and lateral hysteresis loops. This approach was later used by several groups to probe crystallographic orientation and microstructure effects on switching behavior.[36, 37, 38, 39, 40, 41]

Recently, PFM spectroscopy has been extended to an imaging mode using an algorithm for fast (30-100 ms) hysteresis loop measurements developed by Jesse, Baddorf, and Kalinin.[42] In Switching Spectroscopy PFM, hysteresis loops are acquired at each point of the image and analyzed to yield 2D maps of imprint, coercive bias, and work of switching, providing a comprehensive description of the switching behavior of the material at each point.

The progress in experimental methods has stimulated a parallel development of theoretical models to relate PFM hysteresis loop parameters and materials properties. A number of such models are based on the interpretation of phenomenological characteristics of piezoresponse force spectroscopy (PFS) hysteresis loops similar to macroscopic *P*-*E* loops, such as slope, imprint bias, vertical shift, as illustrated in Fig. 1. In particular, the slope of the saturated part of the loop was originally interpreted as electrostriction; later studies have demonstrated the linear electrostatic contribution to the signal plays the dominant role.

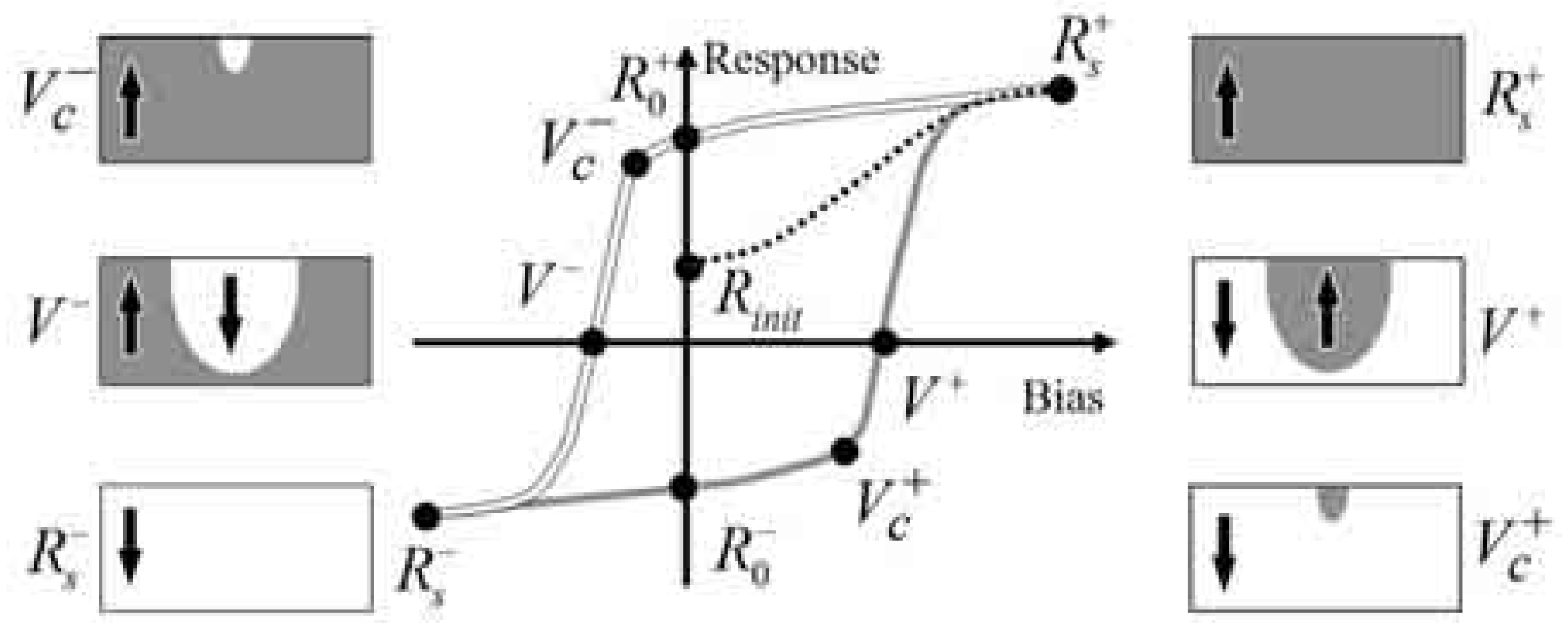

FIG. 1. (Color online) (a) PFM hysteresis loop. Forward and reverse coercive voltages, $V_0^+$ and $V_0^-$, nucleation voltages, $V_{c0}^+$ and $V_{c0}^-$, and forward and reverse saturation and remanent responses, $R_0^+$, $R_0^-$, $R_s^+$, and $R_s^-$, are shown. The work of switching $A_s$ is defined as the area within the loop. The domain structure at the characteristic points of the forward (right) and reverse (left) branches of the hysteresis loop are also shown. Arrows indicate the polarization direction.

Several groups analyzed the effect of non-uniform materials properties, including the presence of regions with non-switchable polarization on parameters such as imprint and vertical shift. In thin films, the vertical shift of the PFM hysteresis loops was interpreted in terms of a non-switchable layer by Saya *et al.*[43] Alexe *et al.*[44] analyzed the hysteresis loop shape in ferroelectric nanocapacitors with top electrodes, obtaining an estimate for the switchable volume of a nanocapacitor. Similar analysis was applied to ferroelectric nanoparticles developed by the self-patterning method[45] by Ma.[46] In all cases, the results were interpreted in terms of ~10 nm of non-switchable layers, presumably at ferroelectric-electrode interface.

A number of authors attempted to relate local PFM hysteresis loops and macroscopic *P-E* measurements, often demonstrating good agreement between the two.[47] This suggests that despite the fundamentally different mechanism in local and macroscopic switching, there may be deep similarities between tip-induced and macroscopic switching processes. A framework for analysis of PFM and macroscopic loops based on Landau theory was developed by Ricinsi et al, [48,49,50] demonstrating an approach to extract local switching characteristics from hysteresis loop shape using first order reversal curve diagrams.

In parallel with tip-induced switching studies, a number of groups combined local detection by PFM with a uniform switching field imposed through the thin top electrode to study polarization switching in ferroelectric capacitor structures. Spatial variability in switching behavior was discovered by Gruverman *et al.* and attributed to strain[51] and flexoelectric[52] effects. In subsequent work, domain nucleation during repetitive switching cycles was shown to be initiated at the same defect regions, indicative of the frozen disorder in ferroelectric structures.[53, 54]

Finally, in a few cases, "abnormal" hysteresis loops having shapes much different then that in Fig. 1 have been reported. Abplanalp *et al.* have attributed the inversion of electromechanical response to the onset of ferroelectroelastic switching.[55] Harnagea has attributed the abnormal contrast to the in-plane switching in ferroelectric nanoparticles.[47,56] Finally, a variety of unusual hysteresis loop shapes including possible Barkhausen jumps and fine structures associated with topographic and structural defects have been observed by Rodriguez *et al.*[57] and Jesse *et al.*[58]

The rapidly growing number of experimental observations and recent developments in PFS instrumentation, data acquisition, and analysis methods requires understanding not only phenomenological, but also quantitative parameters of hysteresis loops, such as numerical value of the coercive bias, the nucleation threshold, etc. Kalinin *et al.*[59] have extended the 1D model by Ganpule *et al.*[60] to describe PFM loop shape in the thermodynamic limit. Kholkin has postulated the existence of nucleation bias from PFM loop observations, in agreement with theoretical studies by Abplanalp,[21] Kalinin *et al.*,[23] Emelyanov,[24] Morozovska and Eliseev.[25] Finally, Jesse *et al.*[42] have analyzed hysteresis loop shape in kinetic and thermodynamic limits for domain formation. However, in all cases, the model was essentially 1D, ignoring the fundamental physics of domain switching. Here, we develop the full 3D model for hysteresis loop formation in PFM including the bias dependence of domain parameters and the relationship between the PFM signal and domain geometry.

### II.2. Domain switching in PFM

The stages of the tip-induced domain growth process during hysteresis loop measurements in a semi-infinite material are domain nucleation, and subsequent forward and lateral domain growth. On reverse bias, both (a) shrinking of the formed domain and (b) nucleation of the domain of opposite polarity are possible.

The analysis of domain dynamics in PFS should qualitatively describe the individual stages in Fig. 2. A number of phenomenological models have been developed based on the classical work of Landauer.[26] In the Landauer model, domain nucleation in ferroelectrics-dielectrics under a *homogeneous* electric field was studied. The model predicted extremely high activation barriers for homogeneous domain nucleation, suggesting the role of defects in polarization switching. This switching behavior is strongly modified in the geometry of a PFM experiment, due to the large (~$10^7$ V/m) electric fields possible in the vicinity of a sharp

AFM probe and the fundamental difference between the uniform field geometry in the Landauer model and the point-contact geometry of PFM.

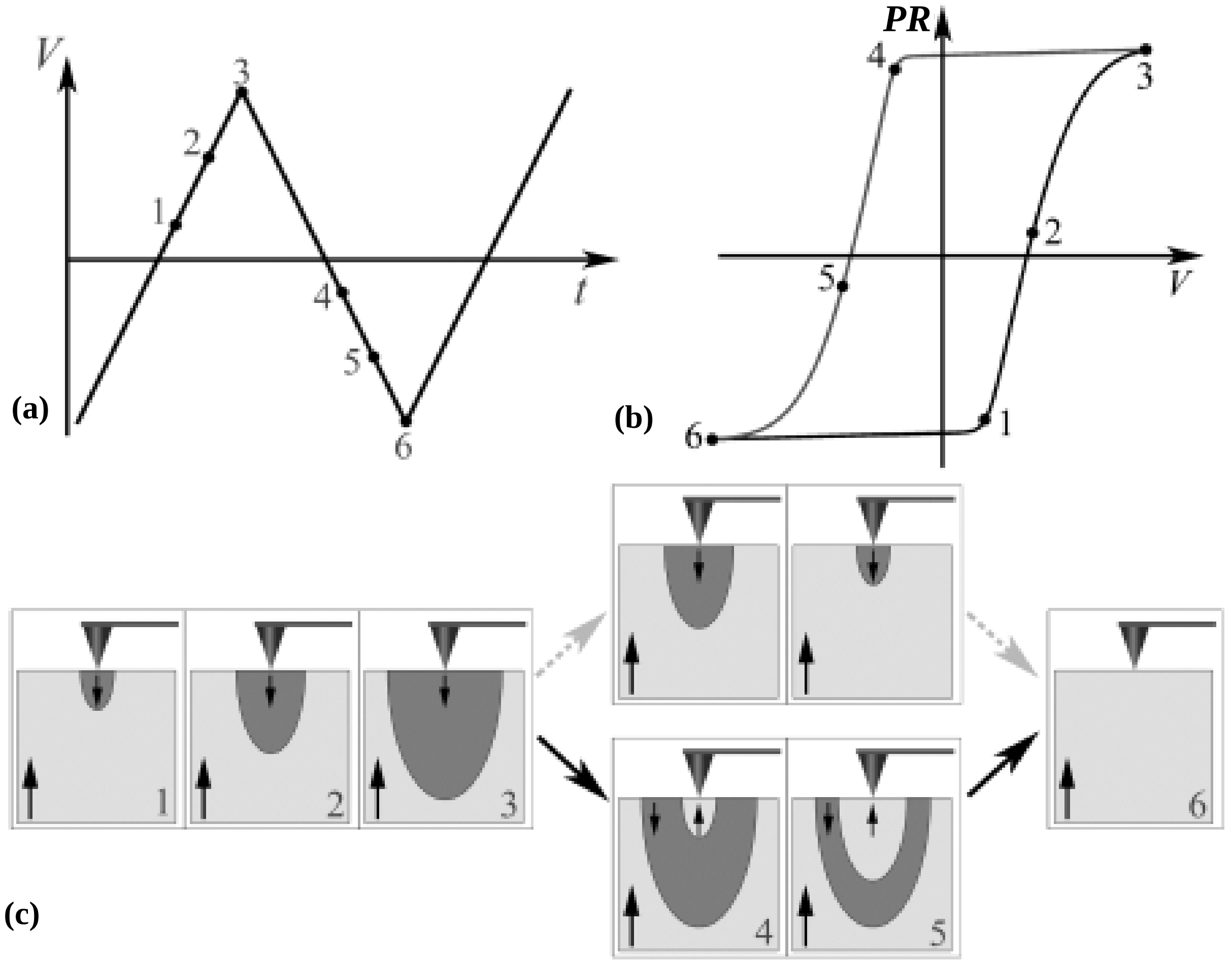

FIG 2. (Color online) Domain evolution with bias dependence for materials with different pinning strengths. (a) Time dependence of voltage and (b) schematics of hysteresis loop. (c) Schematics of the domain growth process. In the purely thermodynamic case (dashed arrows), the domain shrinks with decreased voltage (path 3-4). To account for a realistic loop, the domain size does not change on (3-4) and a domain of opposite polarity nucleates on path 4-6. At point 6, antiparallel domain walls annihilate.

In the original work by Abplanalp,[21] polarization reversal in the *inhomogeneous* electric field of an AFM tip for a semi-ellipsoidal domain with infinitely thin domain walls is considered. The tip was modeled using a point-charge system. In this model, completely uncompensated bound charges on the surface of the ferroelectric were considered and it was shown that the absence of compensation led to an overestimation of the depolarization field energy and thus to a discrepancy between the proposed theory and the available experimental data. It was also predicted that due to the finite charge-surface separation, domain nucleation requires non-zero nucleation bias. This voltage threshold for domain nucleation in the *inhomogeneous* electric field of an AFM tip was then studied by Molotskii,[61] Kalinin *et al.*,[23] Emelianov,[24] and Morozovska and Eliseev[25] using a variety of tip models, as described below.

Using the Landauer model and a point charge approximation for the electric field of an AFM tip, Molotskii[22] obtained elegant closed-form analytical expressions for the domain size dependence on the applied voltage in the case when the surface charges were completely compensated by the external screening charges. The interaction with an external electric field was calculated as if these screening charges were absent. In subsequent work, Molotskii[61] modeled the equilibrium size and kinetics of a cylindrical domain that extended through the film. It was shown that the domain was stable only when the applied voltage exceeded some critical value. For a prolate domain, the depolarization field energy was assumed to be proportional to the film thickness.

Kalinin *et al.*[23] considered the domain nucleation allowing for the electromechanical coupling inside a ferroelectric medium using both the sphere – plane model of the AFM tip and a rigorous solution for the tip-surface indentation problem. This analysis the early stages of the domain growth process as well as higher-order switching phenomena to be studied. It was shown numerically and analytically that the domain nucleation is possible above the threshold value of voltage applied to the tip, i.e. a potential barrier for nucleation exists. Depending on the activation energy, the domain nucleation was classified in terms of first and second order phase transitions.

Similar results were later obtained by Emelyanov,[24] who considered the nucleation of semi-ellipsoidal domains by voltage modulated AFM in ferroelectric films within the framework of classical thermodynamic approach. He analyzed the switching in thin films and classified stages of the switching process and proved that semi-ellipsoidal domains are unstable and transform into cylindrical domains spanning the thickness of the film when reach the bottom electrode.

Recently Morozovska and Eliseev[25] have developed the thermodynamic theory of nanodomain tailoring in thin ferroelectrics films allowing for semiconducting properties, screening, and size effects. The analytical results proved that the nucleation of a cylindrical domain intergrown through the thin film is similar to the first order phase transition. This conclusion completely agrees with recent theoretical predictions and voltage thresholds observed in $Pb(Zr,Ti)O_3$ and $LiTaO_3$ thin films. However, the screening effect on the semi-ellipsoidal domain formation in thicker films and ferroelectric hysteresis and piezoelectric response were not considered.

Following the recent paper by Morozovska and Eliseev,[25] here we extend the thermodynamic theory for hysteresis loop formation in PFM, and analyze the effects of surface conditions and finite conductivity of the material. These results are compared to experimental studies, elucidating role of kinetic effects and pinning on domain formation.

## III. Domain formation and reversal

The driving force for the 180° polarization switching process in ferroelectrics is change in the free energy density.[21,55] The free energy of the nucleating domain is $\Phi = \Phi_U + \Phi_C + \Phi_D$, where the first term is the interaction energy due to coupling between polarization and tip-induced electric field, the second term is the domain wall energy, and the third term is the depolarization field energy. In the Landauer model of switching, the domain shape is approximated as a half ellipsoid with the small and large axis equal to *r* and *l*, correspondingly.

### III.1. Ambient conditions and screening mechanisms

Theoretical descriptions of nanodomain tailoring with local probe under ambient conditions should take into consideration the layer of adsorbed water located below the tip apex,[25, 61] and, more generally, the dynamic and static surface charging and screening phenomena. The static and dynamic properties of charges on ferroelectric surfaces have been recently studied using variable temperature Scanning Surface Potential Microscopy.[62, 63, 64] In a recent study, the role of these charges on polarization dynamics in PFM has been illustrated.[65] In addition to the presence of mobile charges that can redistribute under the action of electric field, a water meniscus appears between the AFM tip apex and a sample surface due to the air humidity. Hence, here we assume that region between the tip apex and domain surface has effective dielectric permittivity, $\varepsilon_e$. Furthermore, the finite conductivity of the medium and electrochemical processes provide additional routes for compensation of surface polarization charge. This screening due to self-ionization and the presence of dissolved ions will be active even in the absence of an applied voltage. Several relevant mechanisms of surface screening can be differentiated:

(a) Screening by ambient charges on the free surface. This process is relatively slow (~10 min) and is limited by the kinetics of the mass-exchange in the vicinity of the sample.[63,64]

(b) Surface charging/electrochemical reactions in the adsorbed water layer at high voltages. The presence of this surface charge on an oxide surface in ambient is a well-known phenomena, as confirmed by charge retention and diffusion on nominally conductive surfaces upon contact electrification or under lateral biasing.[66, 67, 68] In this case, the charging process is relatively fast on the order of milliseconds, as limited by the direct charge transfer from the tip to the surface, surface electrochemical processes, and lateral charge diffusion. Generated charge will form a dipole layer with the polarization charge to minimize coulomb energy, and can potentially increase the apparent area of electrostatic tip-surface contact.

(c) In the absence of tip-surface charge transfer, the tip-induced field effect can result in surface charging. Unlike direct charging, the sign of tip-induced surface charge is opposite to tip bias. This mechanism is relatively rare and can occur if the tip is covered by an oxide.

From this discussion, here we assume that the equilibrium surface charge density $\sigma_S$ can have the form:

$$\begin{cases} \sigma_S = -\mathrm{P_S}, & \text{without screening charges} \\ -\mathrm{P_S} < \sigma_S < \mathrm{P_S}, & \text{partial screening} \\ \sigma_S = +\mathrm{P_S}, & \text{f}\textit{ull}\text{ screening} \end{cases} \tag{1}$$

The relevance of the specific screening mechanism on polarization switching dynamics depends on the relationship between the corresponding relaxation time $\tau_S$ and voltage pulse time $\tau_U$ (i.e. recording time of the domain). “Fast” screening mechanisms with $\tau_S \le \tau_U$ significantly affect the switching process, whereas the "slow" ones with $\tau_S >> \tau_U$ can be ignored. However, these slow mechanisms can significantly affect the domain stability after switching by providing additional channels for minimizing depolarization energy.

### III.2. Free energy functional

To determine the thermodynamic parameters of the switching process including the nucleation bias and equilibrium domain geometry, the domain size is calculated for semi-infinite ferroelectric material using the thermodynamic formalism developed by Morozovska and Eliseev.[69] The Pade approximations for the individual terms in the free energy $\Phi(r,l) = \Phi_S(r,l) + \Phi_U(r,l) + \Phi_D(r,l)$ of the semi-ellipsoidal domain are derived for ferroelectrics-semiconductors allowing for Debye screening and uncompensated surface charges. The relevant calculations and approximations involved are discussed in Appendix A. Below we consider the domain wall surface energy $\Phi_S(r,l)$, the interaction energy with tip-induced electric field, $\Phi_U(r,l)$, and the depolarization field energy $\Phi_D(r,l) = \Phi_{DL}(r,l) + \Phi_{DS}(r,l)$ including the Landauer contribution $\Phi_{DL}(r,l)$ and the depolarization energy $\Phi_{DS}(r,l)$ induced by the surface charges. The latter has not been considered previously,[22, 70] and its inclusion significantly affects the thermodynamic parameters of the switching process.

The domain wall energy $\Phi_S(r,l)$ has the form:

$$\Phi_S(r,l) \approx \frac{\pi^2 \psi_S\, l\, r}{2}\left(1 + \frac{2(r/l)^2}{4 + \pi(r/l)}\right) \tag{2}$$

Pade approximation for the Landauer energy of a semi-ellipsoidal domain including the effects of Debye screening in the material is:

$$\Phi_{DL}(r,l)=\frac{4\pi P_S^2 r^2}{\varepsilon_0\kappa}\frac{R_d n_D(r,l)}{4n_D(r,l)+3R_d(\gamma/l)} \qquad (3a)$$

where

$$n_D(r,l)=\frac{(r\gamma/l)^2}{1-(r\gamma/l)^2}\left(\frac{\operatorname{arcth}\left(\sqrt{1-(r\gamma/l)^2}\right)}{\sqrt{1-(r\gamma/l)^2}}-1\right) \qquad (3b)$$

is the depolarization factor, $\kappa=\sqrt{\varepsilon_a\varepsilon_c}$ is effective dielectric constant of the medium, and, $\gamma=\sqrt{\varepsilon_c/\varepsilon_a}$ is the anisotropy factor.

The energy of the depolarization field created by the surface charges $(\sigma_S-P_S)$ located on the domain face has the form:

$$\Phi_{DS}(r,l)\approx\frac{4\pi r^3 R_d}{\varepsilon_0(16\kappa r+3\pi R_d(\kappa+\varepsilon_e))}\left((\sigma_S-P_S)^2+\frac{2P_S(\sigma_S-P_S)}{1+(l/r\gamma)}\right) \qquad (4)$$

The driving force for the switching process is the tip-surface interaction energy. Here we develop Pade approximation for the interaction energy between a spherical tip and the surface based on image charge series (see Appendix A)

$$\Phi_U(r,l)\approx 4\pi\varepsilon_0\varepsilon_e U R_0\sum_{m=0}^{\infty}q_m\frac{R_d\left((\sigma_S-P_S)F_W(r,0,d-r_m)+2P_S F_W(r,l,d-r_m)\right)}{\varepsilon_0\left((\kappa+\varepsilon_e)R_d+2\kappa\sqrt{(d-r_m)^2+r^2}\right)} \qquad (5)$$

The image charges $q_m$ are located at distances $r_m$ from the spherical tip center, where

$$q_0=1,\quad q_m=\left(\frac{\kappa-\varepsilon_e}{\kappa+\varepsilon_e}\right)^m\frac{\sinh(\theta)}{\sinh((m+1)\theta)}, \qquad (6a)$$

$$r_0=0,\quad r_m=R_0\frac{\sinh(m\theta)}{\sinh((m+1)\theta)},\quad \cosh(\theta)=\frac{d}{R_0}. \qquad (6b)$$

Here $R_0$ is the tip radius of curvature, $\Delta R$ is the distance between the tip apex and sample surface, so $d=R_0+\Delta R$. The function

$$F_W(r,l,z) \approx \frac{r^2}{\sqrt{r^2+z^2}+z+(l/\gamma)} \tag{7}$$

is the Pade approximation of a cumbersome exact expression Eq.(A.3a) obtained originally by Molotskii.[22]

Under the typical condition $\Delta R << R_0$ , i.e. the tip is in contact with the surface, the Eq. (5) can be approximated as:

$$\Phi_U(r,l) \approx \frac{R_d U\, C_t/\varepsilon_0}{(\kappa+\varepsilon_e)R_d + 2\kappa\sqrt{d^2+r^2}}\left(\frac{(\sigma_S - P_S)r^2}{\sqrt{r^2+d^2}+d} + \frac{2P_S r^2}{\sqrt{r^2+d^2}+d \quad +(l/\gamma)}\right), \tag{8}$$

where $d$ is the equivalent charge - surface distance and $C_t$ is the effective tip capacitance. Note, that earlier Molotskii[22] and Morozovska and Eliseev[25] used a capacitance approximation (CA), in which $d = R_0 + \Delta R \approx R_0$ and $C_t \approx 4\pi\varepsilon_0\varepsilon_e R_0 \sum_{m=0}^{\infty}\left(\frac{\kappa-\varepsilon_e}{\kappa+\varepsilon_e}\right)^m \frac{\sinh(\theta)}{\sinh((m+1)\theta)}$ . At the same time, Kalinin *et al.* [71] and Abplanalp[55] used the effective point charge approach, in which the tip is substituted by a single point charge $Q = U \cdot C_t$ at distance $d$ from the surface such that the radius of curvature of equipotential surface in the point of contact is $R_0$ and the potential is equal to $U$. In the particular case $R_d > R_0$, the effective distance $d \approx \varepsilon_e R_0/\kappa$ and $C_t \approx 4\pi\varepsilon_0\varepsilon_e R_0(\kappa+\varepsilon_e)/2\kappa$ respectively. In all other cases, the effective charge-surface separation $d(R_0, R_d)$ should be found numerically (see Appendix A for details). Here we note that the use of the effective point charge model (EPCM) yields a good approximation of the field behavior in the vicinity of the tip surface junction, which is significantly underestimated in the CA model. At the same time, for large separations from contact area, the CA model provides better results.

Hence, the EPCM is better suited to describe nucleation processes, and the CA model is preferential for the description of the latter stages of the switching process. The exact sphere-plane model, while more cumbersome, allows the switching process to be described on all length scales, and is used in this work.

### III.3. The free energy maps and influence of surface screening

The thermodynamics of the switching process can be analyzed from the bias dependence of the free energy of the nascent domain. The dependence of the free energy for a nucleating domain

$$\Phi(r,l) = \Phi_S(r,l) + \Phi_U(r,l) + \Phi_D(r,l) \tag{9}$$

on domain radius, $r$, and length, $l$, can be represented as a free energy surface. The characteristic maxima, minima, and saddle point define the possible stable, unstable, and metastable polarization states.

The evolution of the free energy surface with bias for typical tip and materials parameters is illustrated in Fig. 3. For small biases $U < U_S$, the free energy is a positively defined monotonic function of domain sizes, corresponding to the absence of a stable domain. For biases $U_S < U < U_{cr}$, the local minimum $\Phi_{\min} > 0$ arises, corresponding to a metastable domain with $r_{ms}$ and $l_{ms}$. The corresponding energy is referred to as the equilibrium domain energy $\Phi(r_{ms}, l_{ms}) = E_{ms}$. Finally, for $U \geq U_{cr}$, the absolute minimum $\Phi_{\min} < 0$ is achieved for $r_{eq}$ and $l_{eq}$, corresponding to a thermodynamically stable domain. The value $U_{cr}$ determines the point where the homogeneous polarization distribution becomes absolutely unstable. Such "threshold" domain nucleation is similar to the well-known first order phase transition.

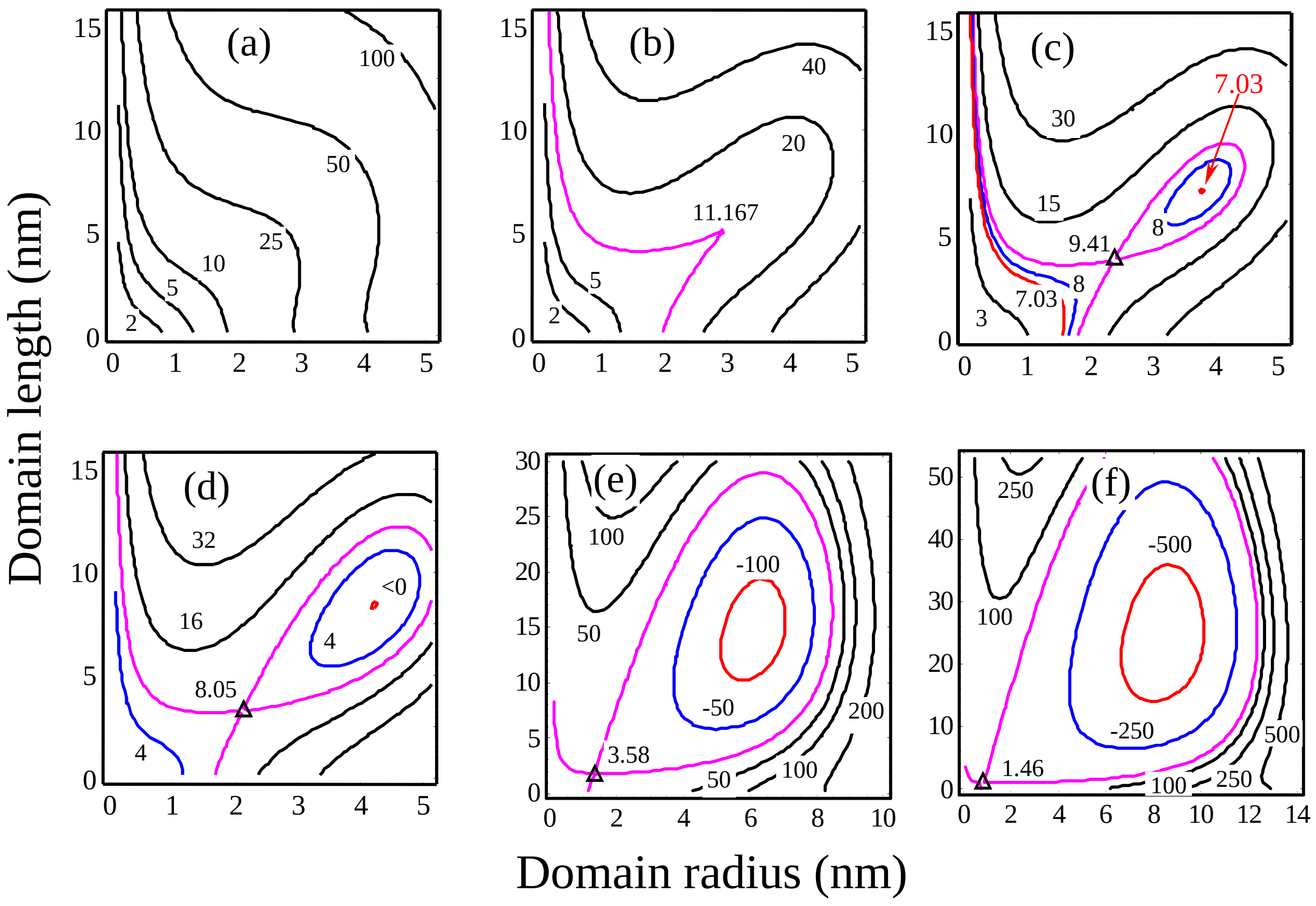

FIG. 3. (Color online) Contour plots of the free energy surface under the voltage increase: (a) domain is absent ($U = 2$ V); (b) instability point - local minimum became to appear ($U = 2.344$ V); (c) saddle point and metastable domain appears ($U = 2.4$ V); (d) transition point - the stable domain appears ($U_{cr} = 2.467$ V); (e, f) stable domains growth ($U$ = 3; 4 V). Figures near the contours are free energy values in eV. Triangles denote saddle point (nuclei sizes). Material parameters: Debye screening radius $R_d = 500\,nm$, $P_S \approx 0.5\,C/m^2$, $\psi_S \approx 150\,mJ/m^2$, $\varepsilon_a \approx 515$, $\varepsilon_c = 500$ correspond to the lead zirconate titante (PZT)6B solid solution and tip-surface characteristics: $\varepsilon_e = 81$, $R_0 = 50nm$, tip touches the sample ($d = 8.1nm$); uncompensated surface charges density $\sigma_S = -P_S$. Note, that the expression Eq. (8) for the EPCM model was used in calculations.

The minimum point (either metastable $\{r_{ms}, l_{ms}\}$ or stable $\{r_{eq}, l_{eq}\}$) and the coordinate origin are separated by the saddle point $\{r_S, l_S\}$. The corresponding energy $\Phi(r_S, l_S) = E_a$ is an activation barrier for domain nucleation, while domain parameters $\{r_S, l_S\}$ represent the critical nucleus size. This behavior is due to the finite value of electric field on the surface, precluding nucleation at small biases.

From Eq. (9) both interaction energy and depolarization energy decrease as a result of surface screening, i.e. with surface charge density changing from $-P_S$ to $+P_S$. Remarkably, the free energy is always positive at $\sigma_S = +P_S$ and $U > 0$, rendering domain nucleation impossible at $\sigma_S \to P_S$, since the positive bulk depolarization energy and domain wall surface energy are independent on $\sigma_S$. This analysis illustrates that surface screening is a necessary condition for domain nucleation in PFM, in agreement with studies by Tagantsev illustrating the role of electrode interface on domain nucleation.[72]

The evolution of the free energy surfaces in Figs. 4 (a - f) illustrates the role of screening on the thermodynamics of domain formation. From the data in Fig. 4, it is clear that domain switching is controlled by screening charge density $\sigma_S$ within the framework of the thermodynamic model. Charge screening ($\sigma_S > +P_S$) results in a decrease of the dragging electrostatic force caused by the charged probe. In the case of full screening ($\sigma_S = +P_S$) the dragging force is absent.

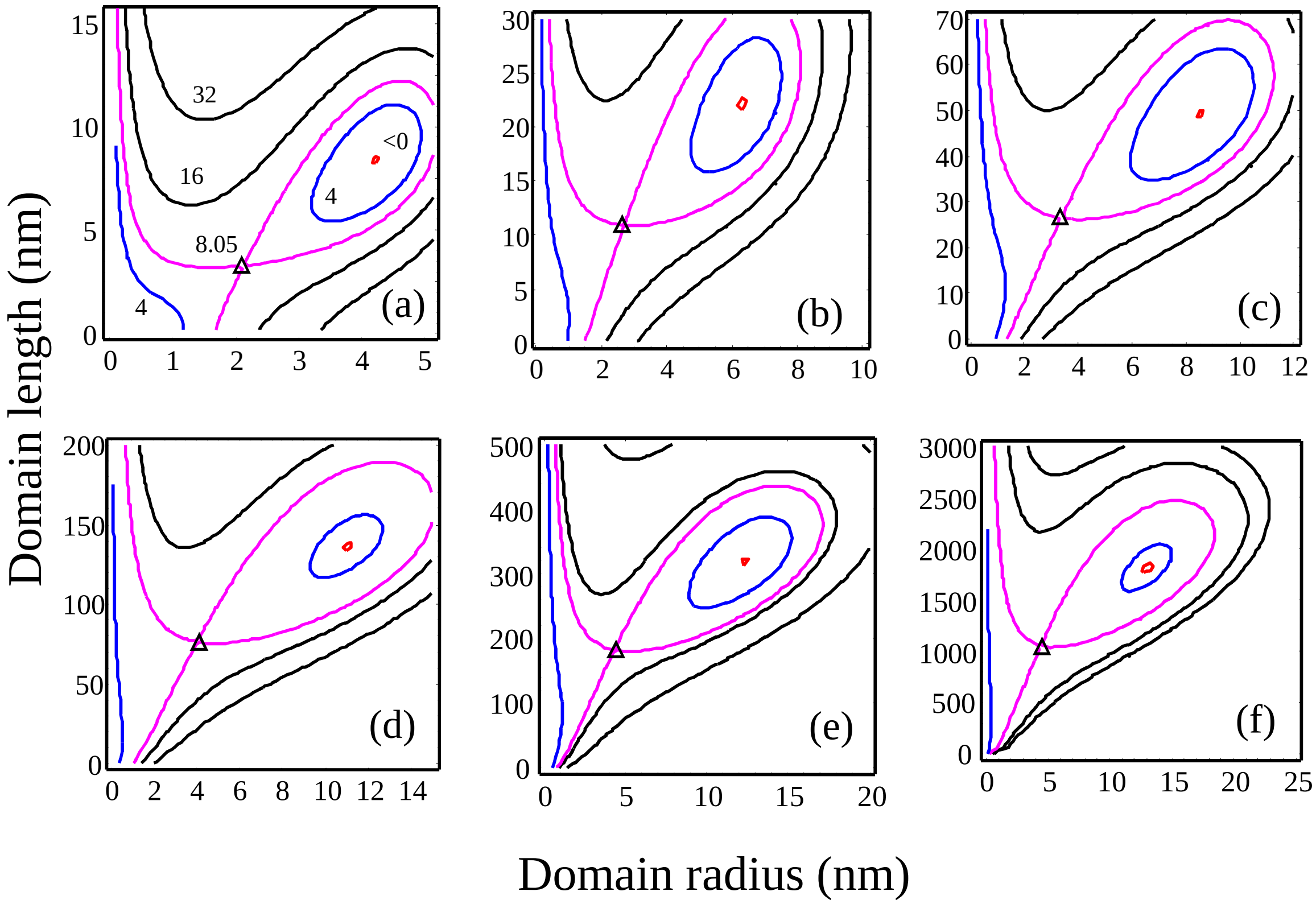

FIG. 4. (Color online) The evolution of free energy map with surface screening charge density. Shown are surfaces for voltages corresponding to onset of thermodynamic switching. (a) $\sigma_S = -P_S$ ($U_{cr} = 2.47$ V); (b) $\sigma_S = -0.5P_S$ ($U_{cr} = 4.21$ V); (c) $\sigma_S = 0$ ($U_{cr} = 9.07$ V); (d) $\sigma_S = +0.5P_S$ ($U_{cr} = 36.75$ V); (e) $\sigma_S = +0.75P_S$ ($U_{cr} = 153.45$ V); (f) $\sigma_S = +0.95P_S$ ($U_{cr} = 4087$ V). All material parameters and designations are given in Fig. 3.

This analysis suggests that environmental effects and surface state will provide critical influence on polarization switching processes in PFM. Variation of imaging medium from ambient to vacuum or inert gas, distilled water, electrolyte and some chemically inert liquid dielectric can illustrate these effects. In particular, the dependence of critical voltage $U_{cr}$ values over ambient conditions (if any) could clarify the surface screening influence. Notably,

Terabe *et al.*[73] have demonstrated that values of $U_{cr}$ on +Z and -Z cuts of $LiNbO_3$ or $LiTaO_3$ crystals differ by a factor of 2, illustrating the effect of surface state on switching mechanism.

### III.4. Bias dependence of energetic and geometric parameters of domains

Activation energy for nucleation and critical domain size can be determined from the saddle point on the $\Phi(r,l)$ surface. Thermodynamic nucleation bias corresponds to the condition when the energy of the minimum on the $\Phi(r,l)$ surface becomes negative, i.e. domain becomes thermodynamically stable. The equilibrium domain size can be determined from the minimum of the $\Phi(r,l)$ surface. Shown in Fig. 5 are the activation energy for nucleation (a), critical nucleus sizes (b,c), domain energy (d) and equilibrium domain sizes (e,f) calculated in the framework of sphere-plane model, modified point-charge model and the CA model (compare dotted, solid and dashed curves in Fig. 5).

Note that the critical domain shape is close to the semi-spherical independently on the adopted model, whereas equilibrium domain is always prolate [compare Fig. 4 (b,c) with Fig. 4 (e,f)]. From Fig. 4, domain formation is impossible below a certain nucleation bias, $U_{cr}$, while above this voltage, the nucleus sizes rapidly decrease with bias.

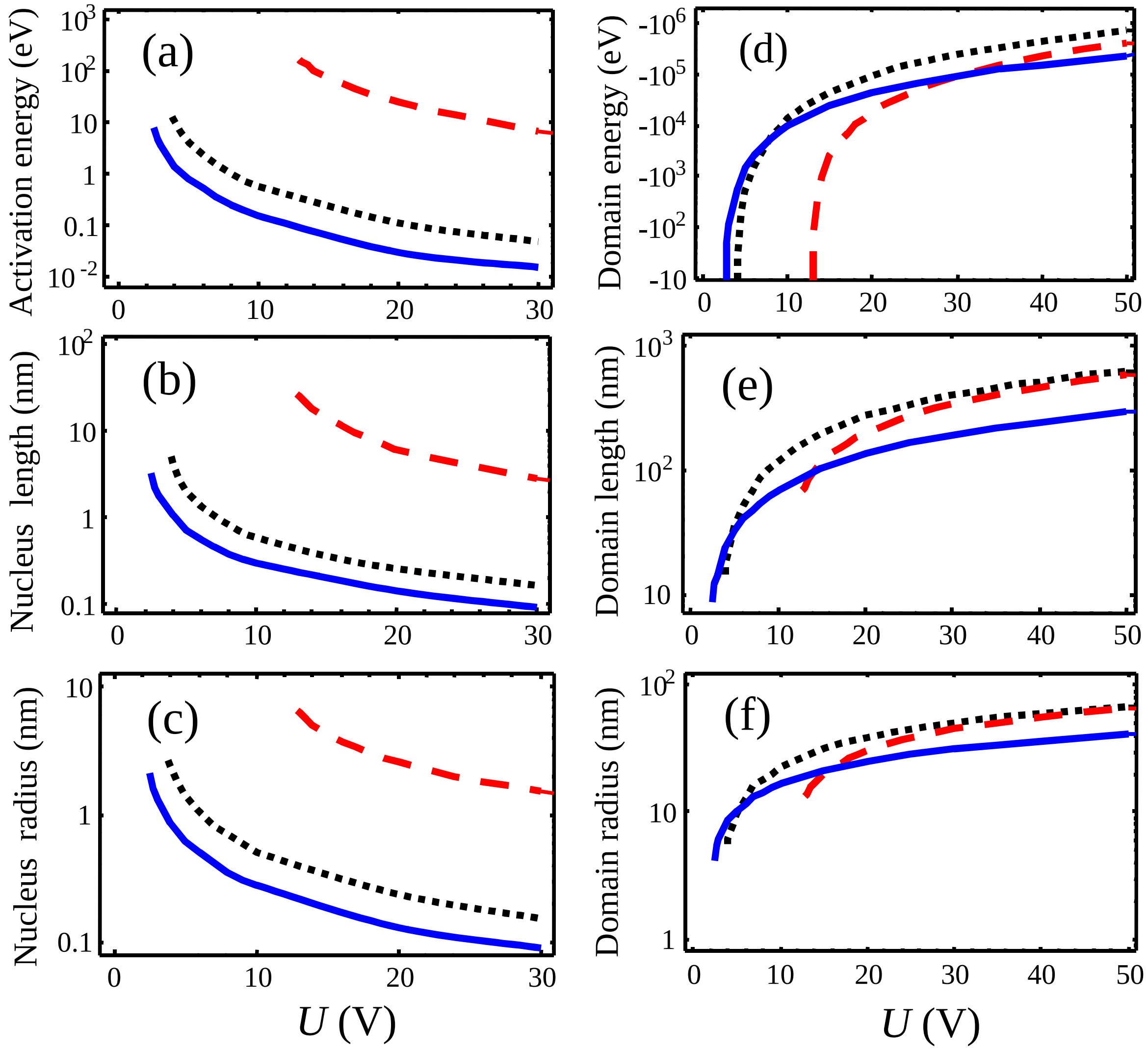

FIG. 5. (Color online) Bias dependence of (a) activation energy for nucleation (eV) and nucleus (b) length and (c) radius, and (d) equilibrium domain energy (eV) and (e) length and (f) radius calculated for PZT6B. Solid curves represent modified point charge approximation of the tip; dotted ones correspond to the exact series for sphere-tip interaction energy, dashed curves represent the capacitance model. Material parameters are given in Fig. 3. Calculations are performed for complete screening, $\sigma_S = -P_S$.

Assuming that the characteristic time for nucleation is $\tau = \tau_0 \exp(E_a / k_B T)$ and the attempt time $\tau_0 = 10^{-13}\, s$, the thermal activation of domain nucleation in the PFM experiment requires an activation barrier below 0.7 - 0.8 eV ($\tau = 10^{-3} - 1$ s) corresponding to the tip voltages $U = 5...8$ V. Interestingly, in the CA model, the activation energy is very high (barrier about 200 eV for $U_{cr} \approx 10$ V) making the thermal activation of the domain nucleation process impossible. The sphere-plane and EPCM developed here allow for a high field concentration at the tip-surface junction, and thus are suited for the description of domain nucleation in PFS. Note that the applicability of the model requires the critical domain size to be larger then several correlation lengths. The correlation length cannot be smaller than several lattice constants, thus the nucleation barrier disappears at $U \geq 15$ V.

Shown in Fig. 6 are the activation energies for nucleation (a) and nucleus sizes (b,c), critical voltage (d) and sizes (e,f) calculated in the framework of the sphere-plane model, the modified point charge model, and the CA model for different screening conditions on the surface. It is clear from the figure that all critical parameters rapidly increase under the charge density $\sigma_S$ increase from $-P_S$ to $+P_S$. Namely, activation energy, critical voltage, nucleus and critical domain length increase faster then exponentially, whereas nucleus and critical domain radiuses increase linearly (modified point charge) or faster then linearly (sphere-plane) depending on the tip model.

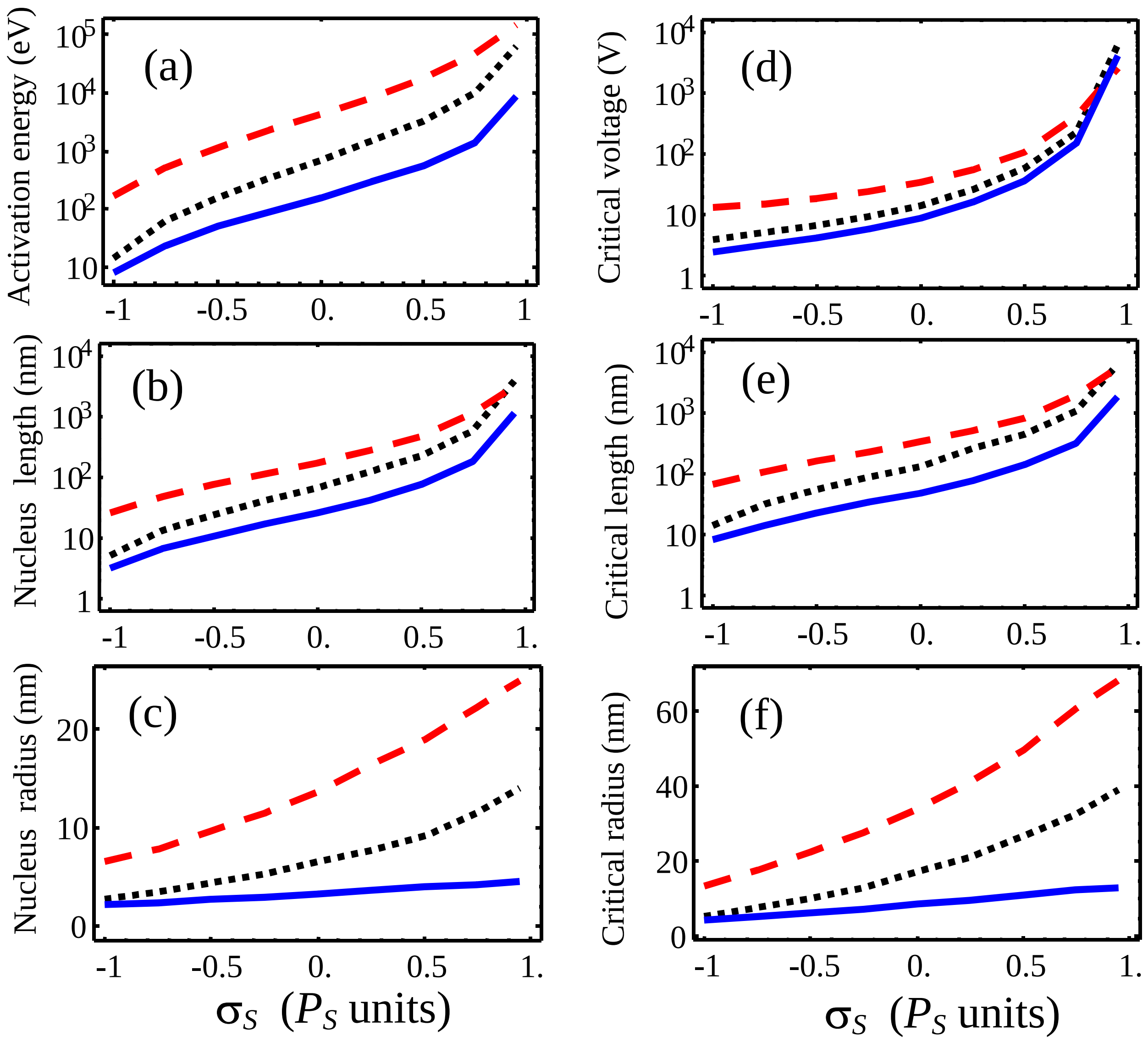

FIG. 6. (Color online) (a) Activation energy (eV) at $U_{cr}$ and nucleus sizes (b,c); (d) critical voltage $U_{cr}$ (V), critical domain sizes (nm): length $l(U_{cr})$ (e) and radius $r(U_{cr})$ (f) vs. surface charge density $\sigma_S$ (in $P_S$ units) calculated for PZT6B. Solid curves correspond to EPCM model of the tip; dotted ones correspond to the exact series for sphere-tip interaction energy, dashed curves represent the CM. Material parameters are given in caption to Fig.3.

In the framework of the model considered here, the activation barrier for nucleation at the onset of domain stability (see Fig. 6a) is minimal for complete screening at $\sigma_S = -P_S$ (10 eV) and increases up to $10^5$ eV for $\sigma_S \to +P_S$.[74] Also, the barrier height strongly decreases with further voltage increase $U > U_{cr}$ at all $\sigma_S$ values. Note, that the barrier calculated in the inhomogeneous electric field of the tip is 3 - 5 orders lower than the one calculated by Landauer for the homogeneous electric field, the values obtained at $\sigma_S > -P_S$ are still too high for thermal fluctuations to cause the domain nucleation at $U \approx U_{cr}$. Thus the observed domains could either originate at higher voltages in the perfect ferroelectric sample (see comments to Fig. 5), or nucleation must be defect-related. Note, that significantly lower barriers correspond to $BaTiO_3$ and Rochelle salt, allowing for the lower values of surface energy ( $\psi_S \approx 5\ mJ/m^2$ and $\psi_S \approx 0.06\, mJ/m^2$ respectively). The overscreening of the surface due to direct tip-surface charge transfer is likely to minimize the activation energy for nucleation further.

### III.5. Ferroelectric hysteresis in thermodynamic limit

Based on the evolution of the free energy surfaces in Fig. 3, the following scenario for hysteresis loop formation in the thermodynamic limit emerges. Below the bias $U_S$, domain formation is thermodynamically impossible. On increasing the bias above $U_S$, the local minimum corresponding to a metastable domain appears. The domain becomes thermodynamically stable above critical bias $U_{cr}$. For an infinitely slow process, domain nucleation below the tip becomes possible at this bias. Realistically, nucleation will proceed at higher bias when the activation energy for nucleation becomes sufficiently low. On subsequent increase of tip bias, the domain size increases. However, due to the 1/*r* decay of electrostatic fields, the domain size always remains finite. On decreasing the bias, the domain becomes metastable at $U_{cr}$ and disappears at $U_S$, resulting in intrinsic thermodynamic hysteresis in hysteresis loop shape.

Description of the thermodynamics and kinetics of domain switching can be greatly facilitated by closed-form analytical expression for the bias dependence of characteristic points on the free energy map. Here, we obtain approximate analytical results within the modified point charge model as proposed in Appendix B. The approximate parametrical dependences for hysteresis curves $U(r)$ valid for $r < d$ have the form:

$$\begin{cases} U(r) \approx \left( \dfrac{3\sqrt{2 f_S} + \sqrt{18 f_S + 8(1-\sigma_S/P_S)\left(3 f_D (1-\sigma_S/P_S)^2 (r/d)^2 - f_S\right)}}{4\sqrt{f_U}(1-\sigma_S/P_S)\sqrt{r/d}} \right)^2 \\ l(r) \approx 2\gamma d\left(\sqrt{\dfrac{2 f_U U}{f_S}}\sqrt{r/d} - 1\right) \end{cases}, \qquad (10)$$

where $f_S = \pi^2 \psi_S d^2 \gamma$, $f_D = \dfrac{4P_S^2 d^3}{3\varepsilon_0(\kappa+\varepsilon_e)}$, $f_U = \dfrac{R_d (C_t/\varepsilon_0) P_S d}{2(\kappa+\varepsilon_e)R_d + 4\kappa d}$, $C_t \approx 4\pi\varepsilon_0\varepsilon_e R_0 \dfrac{\kappa+\varepsilon_e}{2\kappa}$ and $d = \varepsilon_e R_0/\kappa$. Using Eqs. (10) and condition $\Phi(r,l) = 0$, the following approximate expressions for critical voltage $U_{cr}$ and sizes $r(U_{cr})$, $l(U_{cr})$ are derived:

$$U_{cr} = \sqrt{\frac{f_D f_S}{1-\sigma_S/P_S}} \frac{1}{2 f_U} \frac{\left(3 + \sqrt{9 - 4(1-\sigma_S/P_S)}\right)^2}{\sqrt{3 - (1-\sigma_S/P_S) + \sqrt{9 - 4(1-\sigma_S/P_S)}}}, \qquad (11)$$

$$r(U_{cr}) = d\sqrt{\frac{f_S}{f_D(1-\sigma_S/P_S)^3}}\sqrt{3 - (1-\sigma_S/P_S) + \sqrt{9 - 4(1-\sigma_S/P_S)}}, \qquad (12)$$

$$l(U_{cr}) = 2\gamma d\left(\frac{3 + \sqrt{9 - 4(1-\sigma_S/P_S)}}{1-\sigma_S/P_S} - 1\right). \qquad (13)$$

Note, that approximate dependences (11)-(13) cannot reflect the exact behavior of the system at $\sigma_S \to P_S$ allowing for the fact that one could not neglect the Landauer energy (3) at $(1-\sigma_S/P_S)\to 0$. However, one can see that domain nucleation disappears at $\sigma_S \to P_S$ as should be expected, because the interaction energy disappears. Our numerical calculations proved, that the functional dependences (11)-(13) reflect the overall picture rather well (see Fig. 7(a)).

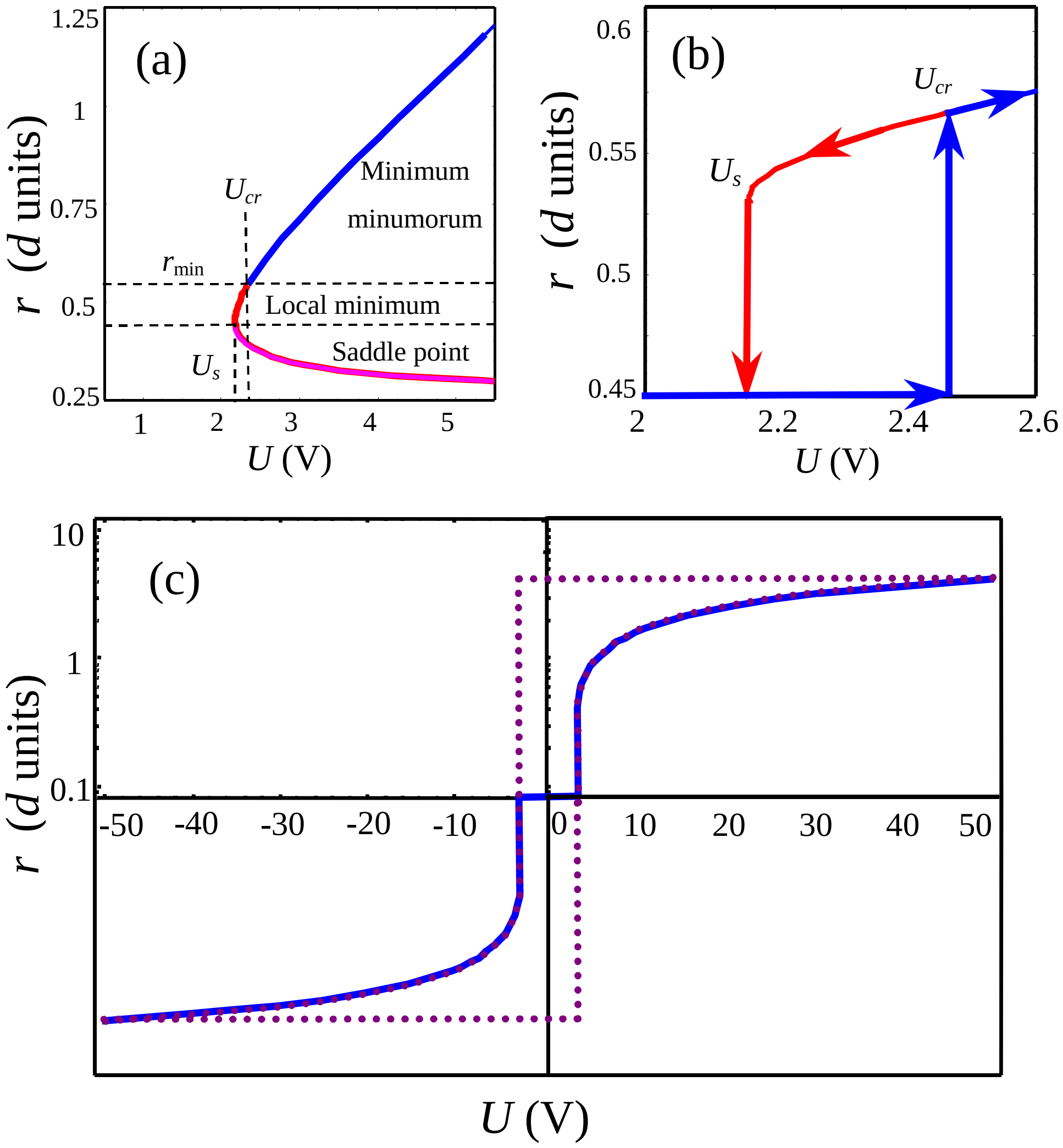

FIG. 7. (Color online) (a) The dependence Eq. (10) of domain radius *r* (in *d* units) vs. voltage *U*. (b) Ferroelectric hysteresis in thermodynamic limit. (c) Full loop in thermodynamic limit

(solid curve) and weak pinning (dotted curve). Here blue curves designate the equilibrium path. All material parameters for PZT6B are same as in Fig. 3.

## IV. Piezoelectric response in final and intermediate states

The analysis in Section III describes the domain evolution with bias. To calculate the shape of the PFM hysteresis loop, the geometric parameters of the domain, i.e. length *l* and radius *r*, must be related to the measured PFM signal. This relationship, once established, will be equally applicable to the thermodynamic theory developed in Section III, the kinetic theory developed by Molotskii and Shvebelman[70, 75], and for data analysis in the PFM experiment.

To establish the relationship between domain parameters and the PFM signal, we utilize the decoupled Green's function theory by Felten *et al.*[76] This approach is based on (1) the calculation of the electric field for rigid dielectric ($d_{ijk} = e_{ijk} = 0$), (2) the calculation of the stress field $X_{ij} = e_{kij} E_k$ in piezoelectric materials, and (3) the calculation of the mechanical displacement field using Green's function for non-piezoelectric elastic body. For transversally isotropic material, the tip-induced electric field can be determined using simple image charge models. For the spherical part of the tip apex, the solution is rigorous, while for the conical part of the tip an approximate line-charge model can be used.[77,78] Here, we develop the solution for an effective charge above the ferroelectric surface, and then extend this theory for an arbitrary point charge distribution.

### IV.1 Piezoresponse in the initial state with finite screening

The potential inside the transversally isotropic dielectric material with finite Debye length produced by the point charge $Q$, at the distance $d$ above the surface, is

$$V_Q(\rho,z)=\frac{Q}{2\pi\varepsilon_0}\int_0^\infty dk\frac{k\,J_0(k\rho)}{\varepsilon_e k+\kappa\sqrt{k^2+R_d^{-2}}}\exp\left(-\sqrt{k^2+R_d^{-2}}\cdot(z/\gamma)-k\,d\right), \qquad (14)$$

where $\rho=\sqrt{x^2+y^2}$ and $z\geq 0$ are radial and vertical coordinate. Note, that

$$V_Q(\rho,z)\approx\begin{cases}\dfrac{Q}{2\pi\varepsilon_0(\kappa+\varepsilon_e)}\dfrac{1}{\sqrt{\rho^2+(z/\gamma+d)^2}}, & \dfrac{R_d}{d}>>1,\\ \dfrac{Q}{2\pi\varepsilon_0\kappa}\dfrac{R_d d\exp(-z/R_d\gamma)}{\sqrt{(\rho^2+d^2)^3}}, & \dfrac{R_d}{d}<<1,\end{cases} \qquad (15)$$

in the limit of rigid dielectric. Within the framework of EPCM model, $d=\varepsilon_e R_0/\kappa$ and $Q=2\pi\varepsilon_0\varepsilon_e R_0 U(\kappa+\varepsilon_e)/\kappa$ at $R_d/d>>1$.

Potential distribution $V_Q(\rho,0)$ on the sample surface $z$ = 0 is shown in Fig. 8 for different Debye lengths, $R_d$, and for PZT6B material parameters. It is clear that the potential vanishes much faster at small $R_d$ values than at $R_d\to\infty$, in particular the proportionality $V_Q(\rho,0)\sim 1/\rho^3$ is valid at $R_d\sim 5...50\,\text{nm}$, whereas $V_Q(\rho,0)\sim 1/\rho$ at $R_d>500\,\text{nm}$ as expected from expansions Eq. (15).

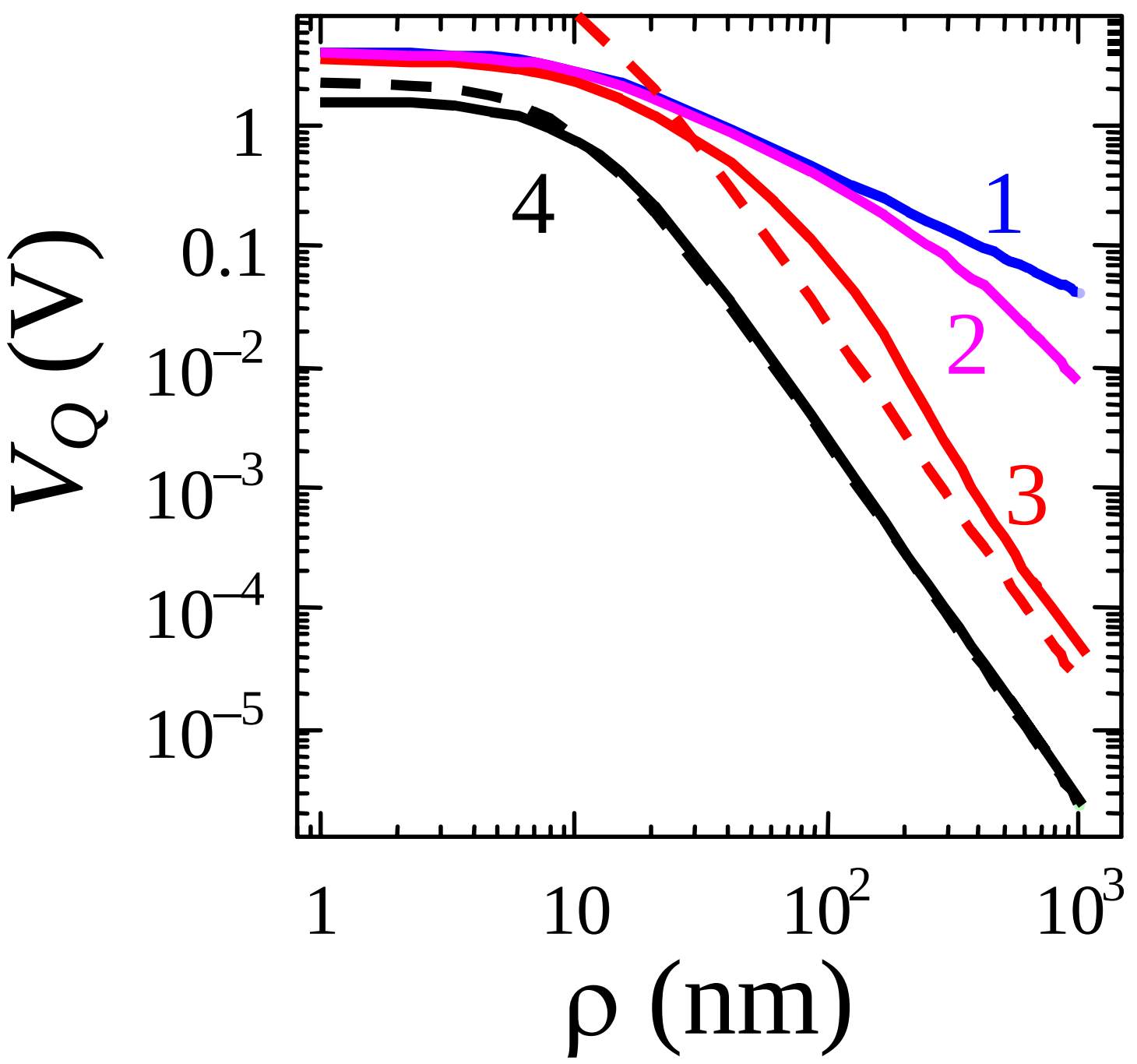

FIG. 8. (Color online) Potential distribution $V_Q(\rho,0)$ calculated from Eq.(14) on the sample surface z=0 at $Q=100\,e$, $d=10\,\text{nm}$ and different $R_d$ radius: ∞ (curve 1), 500 nm (curve 2), 50 nm (curve 3), 5 nm (curve 4). The dashed curves correspond to approximation (15) at $R_d/d<<1$. All material parameters correspond to PZT6B.

The displacement field in the material is calculated using decoupled Green's function approach.[76] The displacement vector $u_3^i(\mathbf{x})$ at position $\mathbf{x}$ is

$$u_3^i(\mathbf{x})=\int_0^{\infty}d\xi_3\int_{-\infty}^{\infty}d\xi_2\int_{-\infty}^{\infty}d\xi_1 e_{kjl}E_k(\xi)\frac{\partial G_{ij}(\mathbf{x},\xi)}{\partial \xi_l}, \qquad (16)$$

where $e_{kjl}=+e_{kjl}$ outside the domain and $e_{kjl}=-e_{kjl}$ inside the domain. In Eq. (16), $\xi$ is the coordinate system related to the material, $e_{kjl}$ are the piezoelectric coefficients ($e_{kij}=d_{klm}c_{lmij}$, where $d_{klm}$ are strain piezoelectric coefficients and $c_{lmij}$ are elastic stiffness) and the Einstein summation convention is used. $E_k(\xi) = -\partial V_Q/\partial x_k$ is the electric field produced by the probe. For typical ferroelectric perovskites, the symmetry of the elastic properties can be approximated as cubic (anisotropy of elastic properties is much smaller then of dielectric and piezoelectric properties) and therefore an isotropic approximation is used.[79] The Green's function for isotropic semi-infinite half-plane is given in Appendix C.

Integration of Eq. (16) for $z = 0, \rho = 0$ over semi-ellipsoidal domain with semi-axes $r$ and $l$ yields the expression for vertical displacement at the position of the tip, i.e. the vertical PFM signal, $u_3^i(0) = u_3(0) - 2\tilde{u}_3(0)$, where $\tilde{u}_3(0)$ is response from semi-ellipsoidal domain and $u_3(0)$ is response from semi-infinite material corresponding to the initial state of the ferroelectric. The displacement in the initial state is:

$$u_3(0) = V_Q(0,0)\left(d_{31} f_1(\gamma, R_d) + d_{15} f_2(\gamma, R_d) + d_{33} f_3(\gamma, R_d)\right). \qquad (17)$$

Functions $f_i(\gamma, R_d)$ depend on the dielectric anisotropy $\gamma$, Poisson ratio $\nu$, screening radius $R_d$ permittivity $\kappa$ and $\varepsilon_e$ (see Fig. 9 for details). Depending on the ratio $\chi^{-1} = 2R_d/d$ we obtained the following approximate expressions (see Appendix C):

$$f_3(\gamma, R_d) \approx -\frac{2\gamma\sqrt{1+\chi^2} + 1 + \chi^2}{\left(\sqrt{1+\chi^2} + \gamma\right)^2}, \qquad (18a)$$

$$f_2(\gamma, R_d) \approx -\frac{\gamma^2}{\left(\sqrt{1+\chi^2} + \gamma\right)^2}, \qquad (18b)$$

$$f_1(\gamma, R_d) \approx -\frac{2\gamma\nu\sqrt{1+\chi^2} + (1+2\nu)\left(1+\chi^2\right)}{\left(\sqrt{1+\chi^2} + \gamma\right)^2}. \qquad (18c)$$

Here, we analyzed specifically the evolution of the response functions with the ratio $R_d/d$. It is clear from Figs. 9 (a,b), that $f_2(\gamma, R_d)$ decreases with decreasing screening length $R_d$ and tends to zero at $R_d/d \to 0$, whereas the other two functions $f_{1,3}(\gamma, R_d)$ increase with decreasing screening length $R_d$ and reach their maximal values at $R_d/d = 0$. The difference is clear, allowing for the fact that $f_{1,3}(\gamma, R_d)$ is caused by the vertical component of the electric field $E_3 = -\partial V_Q/\partial z$, while $f_2(\gamma, R_d)$ originated from the radial one, $E_\rho$.

In addition, effective charge-surface distance *d* depends on $R_d$ value as shown in Fig. 9 (c) (see also Eq.(A.7) in Appendix A). Calculations show that the influence of $R_d$ on the $d$ value is important only for $R_d < R_0$ . Usually $R_d \geq 100\,\text{nm}$ for PZT, whereas tip curvature $R_0 \leq 100\,\text{nm}$. Fig. 9 (d) allows one to choose the region of screening radius $R_d$ where the ratio $R_d/d$ is smaller or greater than unity, and thus which of expressions (18) are valid.

The displacement $u_3(0)$ does not become zero at $R_d/d \to 0$ since the $E_3 = \dfrac{Q}{2\pi\varepsilon_0\kappa\gamma d^2} \neq 0$ in accordance with Eq. (14), however the equipotential surface normal is directed strictly along *z* axes (i.e. $E_\rho \to 0$ at $R_d/d \to 0$). The functions $f_i(\gamma, R_d)$ saturates at $R_d/d \to \infty$. The saturation values $f_i(\gamma)$ correspond to the case of perfect dielectric and are given by Morozovska *et al.*[80] Eqs. (18) define the contributions of different piezoelectric constant to PFM response in the initial and final states of switching process.[79, 81]

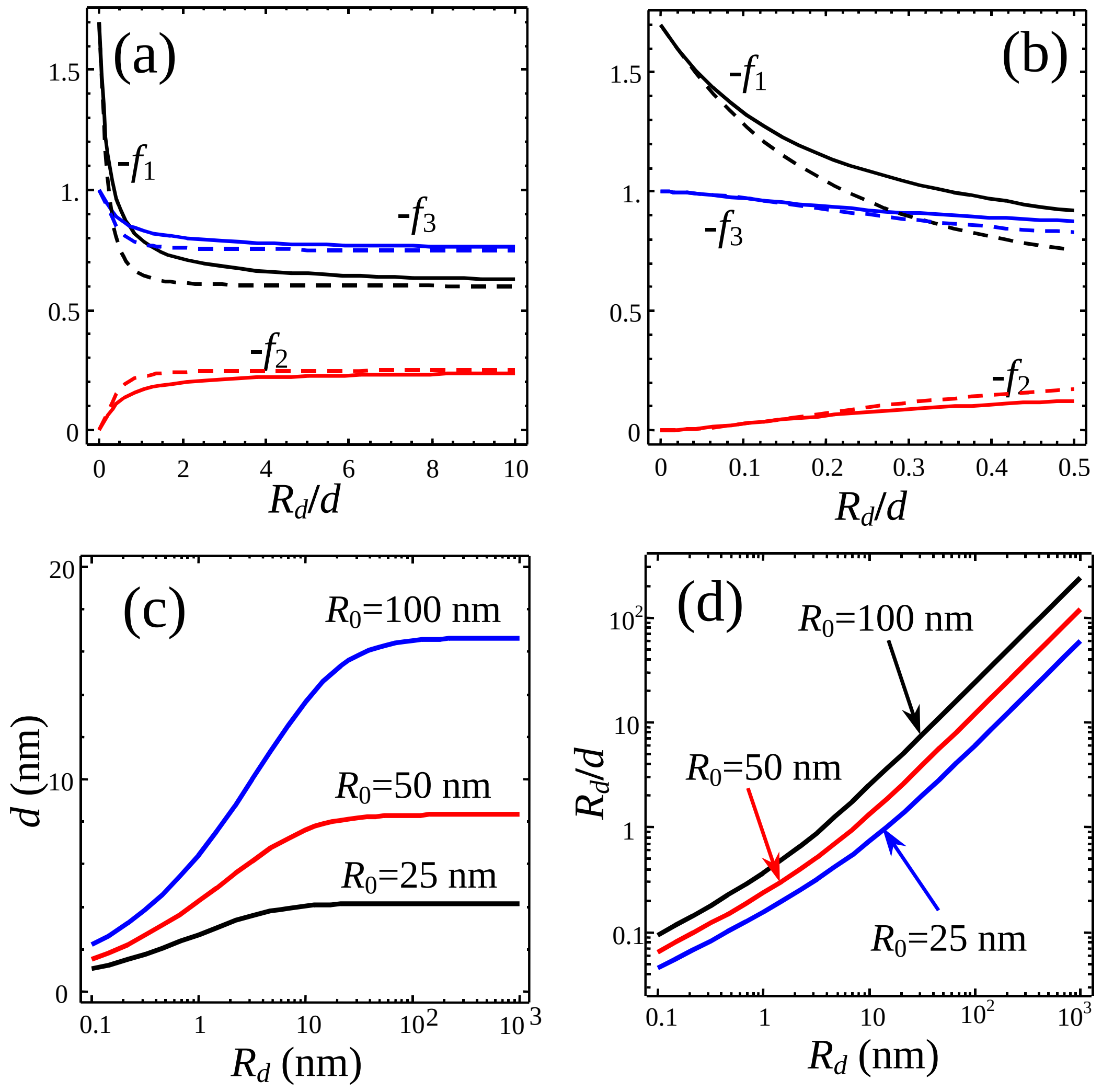

FIG. 9. (Color online) (a,b) Initial states functions $f_i(\gamma, R_d)$ vs. the ratio $R_d/d$ (exact expressions – solid curves, approximations (18) – dashed ones). Effective distance $d$ (c) and ratio $R_d/d$ (d) vs. the screening radius $R_d$ for different tip curvatures $R_0$. Material parameters $\nu = 0.35$, $\gamma = 1$, $\varepsilon_e = 81$ and $\kappa = 500$ correspond to PZT6B.

The PFM response $d_{33}^{eff} = u_3(0)/V_Q(0,0)$ vs. the ratio $R_d/d$ is presented in Fig. 10. Note that the finite Debye length of the material, i.e. the conductivity, reduces the electromechanical response. However, the response does not become zero at $R_d/d \to 0$, due to the finiteness of the electric field

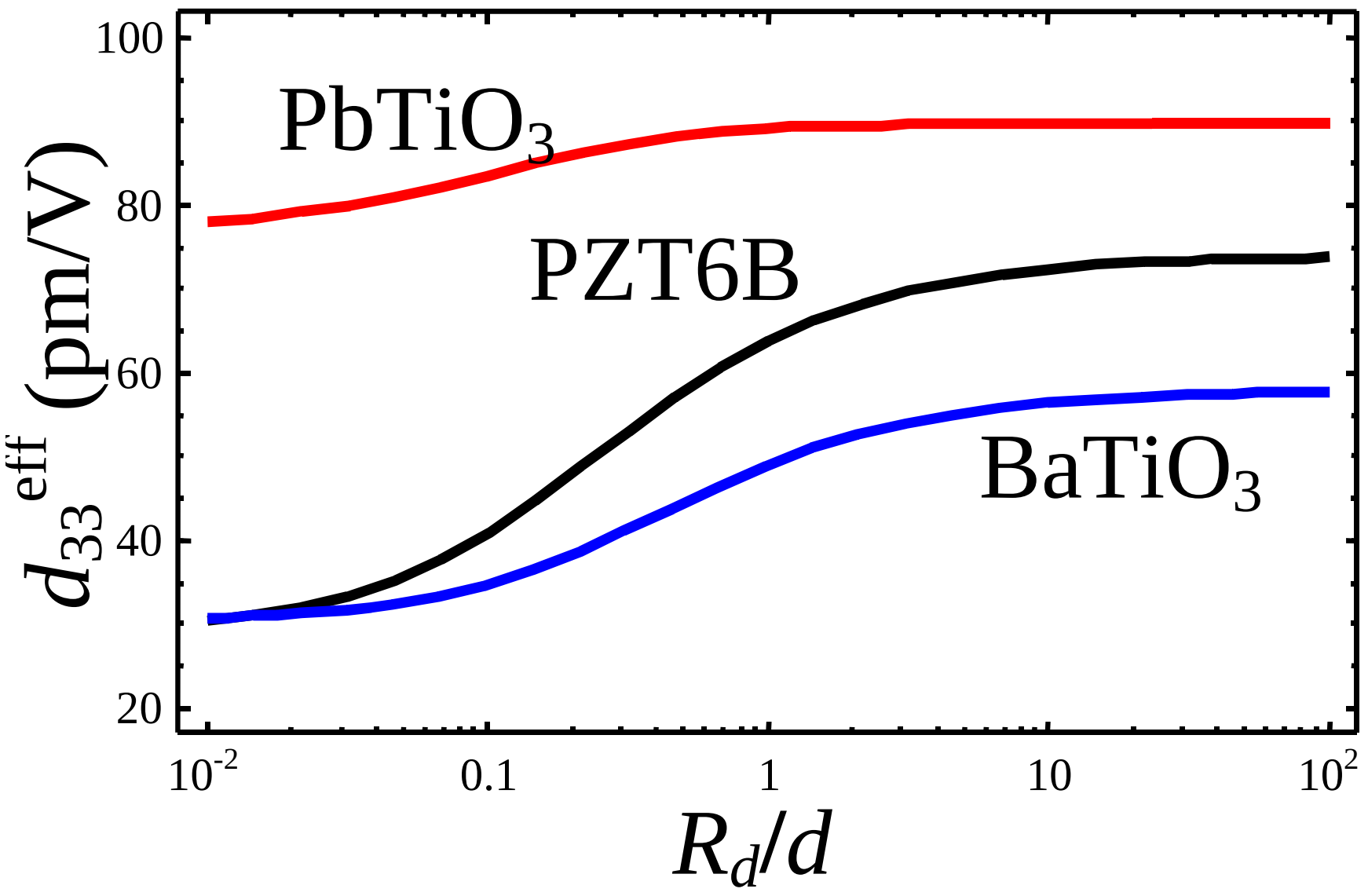

FIG. 10. (Color online) PFM response of the initial state $u_3(0)/V_Q(0,0)$ vs. the ratio $R_d/d$ for $\nu = 0.35$, $\gamma = 1$, $\varepsilon_e = 81$ for ferroelectrics $PbTiO_3$, PZT6B and $BaTiO_3$.

### IV.2. Piezoresponse in the intermediate states

For the intermediate state of the switching process, Eq. (16) acquires the form:

$$u_3^i = V_Q(0,0)\left(d_{31}g_1(\gamma, R_d, r, l) + d_{15}g_2(\gamma, R_d, r, l) + d_{33}g_3(\gamma, R_d, r, l)\right), \qquad (19a)$$

$$g_i(\gamma, R_d, r, l) = f_i(\gamma, R_d) - 2w_i(\gamma, R_d, r, l) \qquad (19b)$$

Functions $w_i = 0$ in the initial and $w_i = f_i$ in the final state of the switching process. For perfect dielectric $R_d \to \infty$, the functions $w_i(\gamma, \nu, R_d, r, l) \equiv w_i^{\infty}(r, l, d)$ are dependent not only on $\gamma$ and $\nu$ but mainly on the domain sizes r, *l* and charge-surface separation *d*. They can be reduced to the one-fold integral representations:

$$w_3^\infty(r,l,d) = -3\int_0^{\pi/2} d\theta \cos^3\theta \sin\theta \frac{R_{dw}(\theta,r,l)}{R_G(\theta,r,l,d)}, \tag{20a}$$

$$w_2^\infty(r,l,d) = \int_0^{\pi/2} d\theta \left(\frac{\gamma d + \cos\theta R_{dw}(\theta,r,l)}{R_G(\theta,r,l,d)} - 1\right) 3\cos^2\theta \cdot \sin\theta, \tag{20b}$$

$$w_1^\infty(r,l,d) = \int_0^{\pi/2} d\theta \left(3\cos^2\theta - 2(1+\nu)\right)\cos\theta\sin\theta \frac{R_{dw}(\theta,r,l)}{R_G(\theta,r,l,d)}, \tag{20c}$$

where the radius $R_{dw}(\theta,r,l) = \dfrac{rl}{\sqrt{r^2\cos^2\theta + l^2\sin^2\theta}}$ determines the ellipsoidal domain wall shape and $R_G(\theta,r,l,d) = \sqrt{(\gamma d + \cos\theta R_{dw}(\theta,r,l))^2 + \gamma^2 \sin^2\theta R_{dw}^2(\theta,r,l)}$. Note, that Eqs. (20) can be extended to arbitrary rotationally invariant domain geometries, e.g. cylindrical or conic, as determined by the functional form of $R_{dw}(\theta,r,l)$.

In the particular case of a cylindrical domain, the functions $w_i^\infty(r,d)$ can be approximated by the following expansions depending on the ratio $\eta = r/d$ (see Fig. 11 and Ref. [80] for details):

$$w_3^\infty(r,d) = \begin{cases} -\dfrac{1+2\gamma}{(1+\gamma)^2} + B_3(\gamma)\dfrac{d}{r}, & r >> d \\ -\dfrac{2}{\gamma}\dfrac{r}{d}, & r << d \end{cases} \approx -\frac{1+2\gamma}{(1+\gamma)^2}\frac{\eta}{\eta + B_3(\gamma)\dfrac{(1+\gamma)^2}{1+2\gamma}} \tag{21a}$$

$$w_2^\infty(r,d) = \begin{cases} -\dfrac{\gamma^2}{(1+\gamma)^2} + B_2(\gamma)\dfrac{d}{r}, & r >> d \\ -\dfrac{1}{2}\left(\dfrac{r}{d}\right)^2, & r << d \end{cases} \approx \frac{-\dfrac{\gamma^2}{(1+\gamma)^2}\eta^2}{\dfrac{\gamma^2}{(1+\gamma)^2}(2 + B_2(\gamma)\eta) + \eta^2} \tag{21b}$$

$$w_1^\infty(r,d)=\begin{cases}-\dfrac{2\nu}{\gamma}\dfrac{r}{d}, & r<<d,\\ -w_3^\infty(r,d)-2(1+\nu)\left(\dfrac{1}{1+\gamma}-B_1(\gamma)\dfrac{d}{r}\right), & r>>d\end{cases}\approx -w_3^\infty(r,d)-\frac{\dfrac{2(1+\nu)}{1+\gamma}\eta}{\eta+B_1(\gamma)(1+\gamma)}$$

(21c)

Constants $B_i(\gamma)$ depend solely on the dielectric anisotropy of material, namely:

$B_1(\gamma)=\dfrac{\pi}{16\gamma^2}\,{}_2F_1\left(\dfrac{3}{2},\dfrac{3}{2};3;1-\dfrac{1}{\gamma^2}\right)$ , $B_3(\gamma)=\dfrac{3\pi}{32\gamma^2}\cdot{}_2F_1\left(\dfrac{3}{2},\dfrac{5}{2};4;1-\dfrac{1}{\gamma^2}\right)$ and

$B_2(\gamma)=\dfrac{3\pi}{32\gamma^2}\left[2\gamma^2\,{}_2F_1\left(\dfrac{1}{2},\dfrac{3}{2};3;1-\dfrac{1}{\gamma^2}\right)-{}_2F_1\left(\dfrac{3}{2},\dfrac{3}{2};4;1-\dfrac{1}{\gamma^2}\right)\right]$, in particular $B_1(1)=\dfrac{\pi}{16}$, and

$B_3(1)=B_2(1)=\dfrac{3\pi}{32}$. Here ${}_2F_1(p,q;r;s)$ is the hypergeometric function.

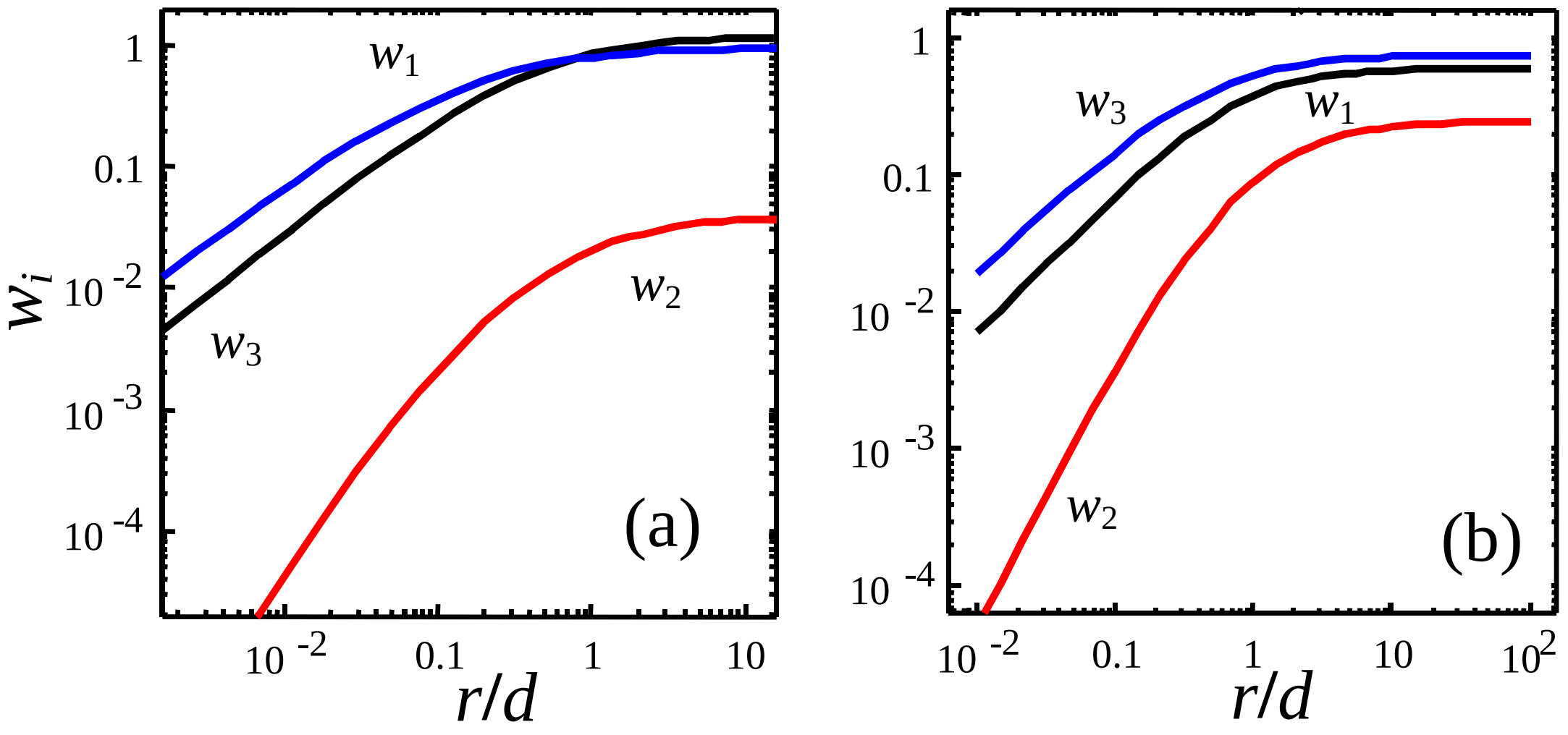

FIG. 11. (Color online). Functions $w_i^\infty(r,d)$ at $\nu=0.35$, $\gamma=0.25$ (a), $\gamma=1$ (b).

For materials with dielectric anisotropy $\gamma=1$ like PZT6B, the effective piezoresponse $d_{33}^{eff}=u_3/U$ for EPCM model can be fitted as:

$$d_{33}^{eff} \approx -\frac{3}{4}\left(d_{33} + \left(\frac{1}{3} + \frac{4}{3}\nu\right)d_{31}\right)\frac{\pi d - 8r}{\pi d + 8r} - \frac{d_{15}}{4}\frac{3\pi d - 8r}{3\pi d + 8r} \ . \tag{22}$$

For more complex tip geometries, Eqs. (19) could be summed over corresponding image charge series. In particular, in the case of a spherical tip that touches the surface of perfect dielectric Eq. (19) should be substituted by series on image charges, namely:

$$u_3^i = \frac{Q}{2\pi\varepsilon_0(\varepsilon_e + \kappa)}\sum_{m=0}^{\infty}\frac{q_m}{d_m}\left(d_{31}g_1(\gamma, r, l, d_m) + d_{15}g_2(\gamma, r, l, d_m) + d_{33}g_3(\gamma, r, l, d_m)\right)$$

$$g_i(\gamma, r, l, d_m) = f_i^{\infty}(\gamma) - 2w_i^{\infty}(r, l, d_m), \tag{23}$$

where $Q = 4\pi\varepsilon_0\varepsilon_e U R_0$, $\quad q_m = \left(\frac{\kappa - \varepsilon_e}{\kappa + \varepsilon_e}\right)^m \frac{1}{m+1}$, $\quad d_m = \frac{R_0}{(m+1)}$ .

### IV.3. Effect of Debye screening on piezoresponse in the intermediate states

For ferroelectric-semiconductor with finite Debye screening radius $R_d$, the functions $w_i(\gamma, R_d, r, l)$ have the form of extremely cumbersome irreducible three-fold integrals. In the general case they should be evaluated numerically.

In the particular case of a cylindrical domain shape or a prolate semiellipsoid ($r << l$) the functions $w_i$ are almost independent on the domain length and are given by the following two-fold integrals (see Appendix C), which could be approximated as following ($\lambda^{-1} = R_d / r$):

$$w_3(\gamma, R_d, r) \approx \frac{f_3(\gamma, R_d)\sqrt{1 + \lambda^2}\left(\frac{r}{d}\right)}{\sqrt{1 + \lambda^2}\left(\frac{r}{d}\right) + \frac{B_3(\gamma)}{\left|f_3(\gamma, R_d)\right|}} \sim \begin{cases} \left(f_3(\gamma, R_d) + B_3(\gamma)\frac{d}{r}\right), & R_d \to \infty \\ f_3(\gamma, R_d) + B_3(\gamma)\frac{R_d d}{r^2}, & \frac{r}{R_d} \geq 1 \end{cases} \tag{24a}$$

$$w_2(\gamma, R_d, r) \approx \frac{f_2(\gamma, R_d)\sqrt{\left(1+\lambda^2\right)^3}\left(\frac{r}{d}\right)}{\sqrt{\left(1+\lambda^2\right)^3}\left(\frac{r}{d}\right) + \frac{B_2(\gamma)}{\left|f_2(\gamma, R_d)\right|}} \sim \begin{cases} \left(f_2(\gamma, R_d) + B_2(\gamma)\frac{d}{r}\right), & R_d \to \infty \\ f_2(\gamma, R_d) + B_2(\gamma)\frac{R_d^3 d}{r^4}, & \frac{r}{R_d} \geq 1 \end{cases} \quad (24b)$$

$$w_1(\gamma, R_d, r) \approx (1+2\nu) w_3(\gamma, R_d, r) - \frac{2(1+\nu)}{\gamma}\sqrt{1+\left(\frac{d}{2R_d}\right)^2}\, w_2(\gamma, R_d, r) \quad (24c)$$

From Eqs. (24) the response in the intermediate states saturates as $d/r$ only at $R_d \to \infty$, whereas for finite Debye radii the saturation is faster and scales as $R_d d/r^2$.

Effective piezoresponse in the intermediate state of the switching process and the corresponding domain radius and length voltage dependence are shown in Fig. 12 for different $R_d$ radius. It is clear from Figs. 12 (b,c) that the domain sizes decrease with decreasing $R_d$. Despite the decrease of domain sizes the piezoresponse saturates much more quickly at small values $R_d \leq 10\,\text{nm}$ than at large ones $R_d \geq 10^3\,\text{nm}$.

The reason for the faster response saturation is that the tip potential quickly vanishes for small $R_d$ (see Fig. 8 and Eq.(15)) and leads to a strong decrease of the PFM response region: both surface radius $R_{\max}$ and penetration depth $h_{\max}$ allowing for the facts that the probe electric field $E_z(0,z) \sim \exp(-z/R_d\gamma)$ in accordance with exponential law and $E_z(\rho,0) \sim \left(\rho^2 + d^2\right)^{-3/2}$ in accordance with power law in comparison with the dependence $E_z(\rho,z) \sim (z/\gamma + d)/\left(\rho^2 + (z/\gamma + d)^2\right)^{-3/2}$ valid for perfect dielectric $R_d \to \infty$. The space outside the region ($R_{\max}$, $h_{\max}$) is invisible to PFM, so when the domain radius reaches $R_{\max}$ and height acquires $h_{\max}$ respectively, the response almost saturates.

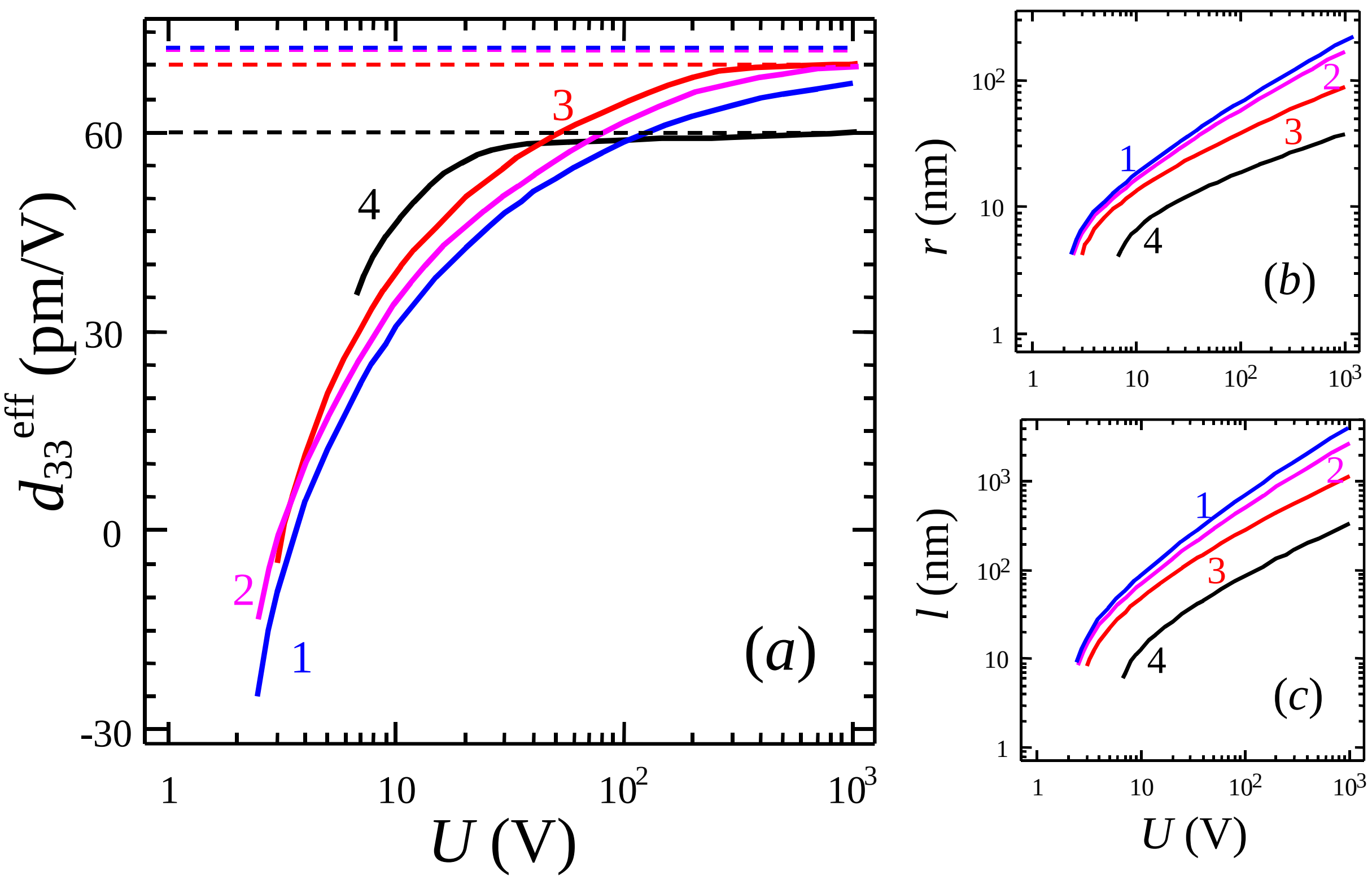

FIG. 12. (Color online) Effective piezoresponse (a) in the intermediate state of switching process and corresponding domain radius (b) and length (c) voltage dependence at different $R_d$ radius: ∞ (curve 1), 500 nm (curve 2), 50 nm (curve 3), 5 nm (curve 4). Dashed curves denote piezoresponse saturation values. Material parameters: $R_0 = 50\,\text{nm}$, $\varepsilon_e = 81$ and $\kappa = 500$ correspond to PZT6B.

Note, that materials such as nearly stoichiometric $BiFeO_3$, $LiNbO_3$ or $LiTaO_3$ typically posses $R_d \geq 10\,\mu\text{m}$ and even slightly doped $BaTiO_3$ has $R_d \sim 100\,\text{nm}$. Thus the Debye screening effect on the nanodomain nucleation and early stages of radial growth is expected to be relatively weak. However, it will significantly affect the vertical domain growth (since $l >> R_d$ is possible) and lateral size saturation at high voltages, resulting in self-limiting behavior due to tip field screening.

## V. Results and Discussion

### V.1. Representative Experimental Results

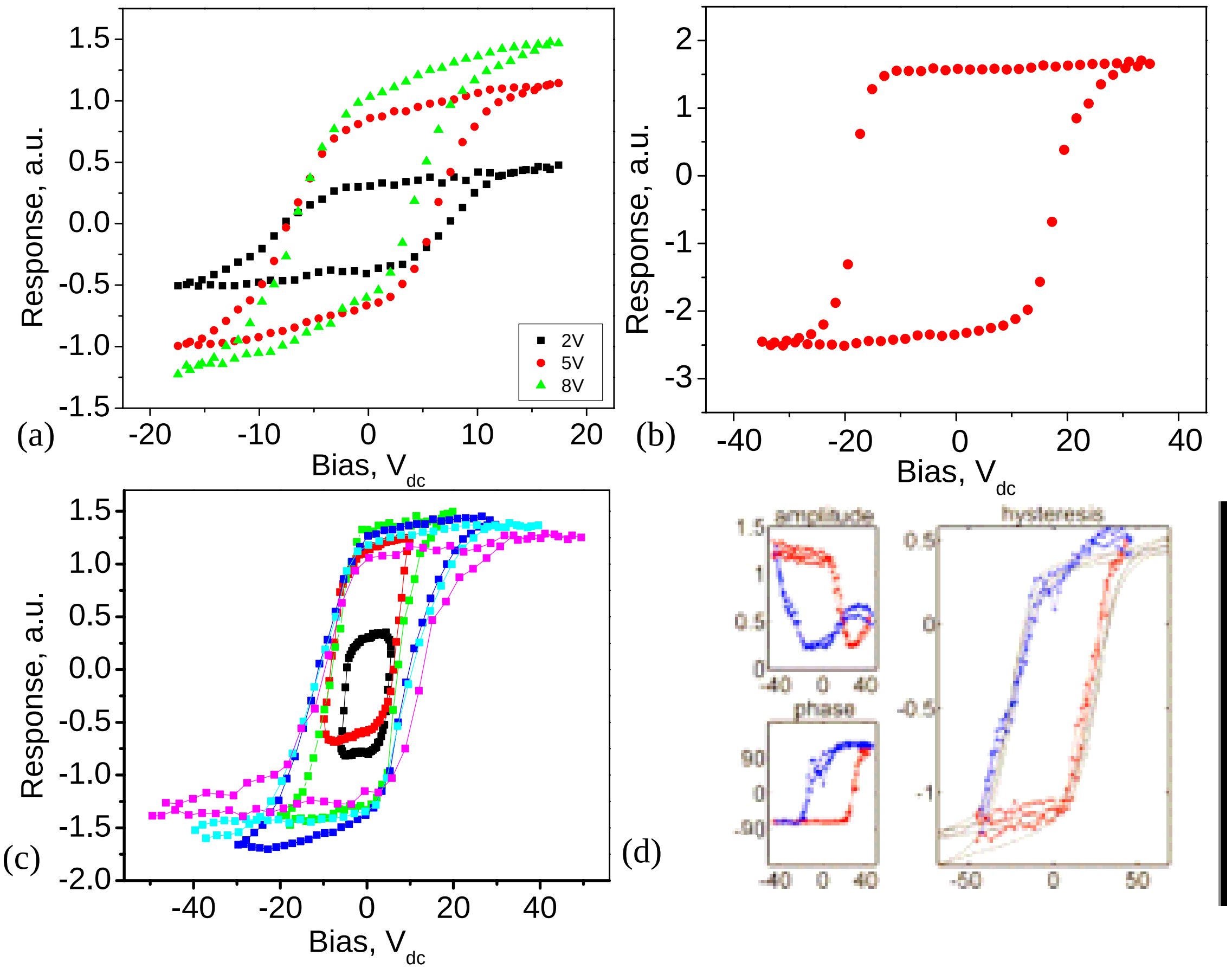

FIG. 13. (Color online) Examples of typical hysteresis loops obtained for several materials. (a) Hysteresis loops obtained on (a) sol-gel PZT film and (b) epitaxial $BiFeO_3$ [sample courtesy of Z. Tong and R. Ramesh, UC Berkeley]. (c) Bias window dependent hysteresis loops on epitaxial PZT thin films [reprinted with permission from Ref. 58]. (d) Hysteresis loops on polycrystalline PZT ceramics [reprinted with permission from Ref. 58].

Shown in Fig. 13 are typical hysteresis loops obtained from a variety of ferroelectric thin film and ceramic samples. In Fig. 13 (a) several loops as a function of ac bias (500 kHz) from a sputtered PZT thin film are shown. Note the presence of hysteretic (forward and reverse branches are different) and saturated (forward and reverse branches saturate) parts of the loop. A saturated hysteresis loop from a multiferroic bismuth ferrite thin film is shown in Fig. 13 (b). The nucleation event is clearly visible. Fig. 13 (c) illustrates the evolution of the PFM hysteresis loop on epitaxial PZT film with bias window. Below the nucleation bias, the switching does not proceed, while above nucleation bias the loop opens up. Finally, shown in Fig. 13 (d) is the hysteresis loop obtained on a polycrystalline PZT ceramic. Note that in most cases, the bias required for nucleation is of the order of 5-10 V.

The experimental loop shape is markedly different than the purely thermodynamic loop in Fig. 7, for which hysteresis is possible only due to the metastability of a bias-induced domain. To account for the observed behavior, we note that the local electrostatic fields are much smaller at the outer domain wall then in the vicinity of the tip-surface junction, and domain wall pinning at the surface, interfaces, defects, and lattice can be sufficiently strong to stabilize the domain. Hence, on reversing the bias, the outer domain wall does not move and reverse switching proceeds thorough the nucleation of a domain of opposite polarity. We refer to the scenario in which the domain size closely follows the thermodynamic model on forward bias, and domain wall does not move on reverse bias, as weak pinning.

The other limiting case is the strong pinning, or kinetic regime, when the domain size is significantly smaller than the thermodynamic prediction and is limited by the bias-dependent mobility of the domain wall. In this case, the hysteresis loop will be significantly broadened compared to the weak pinning case. However, even in this case the thermodynamically determined equilibrium domain size is required to determine the effective driving force for domain wall motion.

### V.2. Modelling loop shape in weakly pinned limit

In the section we analyze the shape of piezoresponse loop for PZT in the weak pinning limit. To calculate the thermodynamic hysteresis loop shape from the bias dependence of the domain size, we assume that the domain evolution follows the equilibrium domain size on the forward branch of the hysteresis loop [see Fig. 7 (c)]. Corresponding piezoelectric loops calculated using thermodynamic parameters derived in Section III using formulae in Section IV are shown in Figs. 14-15.

The initial domain nucleation occurs at $U \geq U_{cr}$ (path 12). Then domain sizes increase under the further voltage increase (path 23). On the reverse branch of the hysteresis loop, the domain does not shrink. Rather, the domain wall is pinned by the lattice and defect (path 34).[82]. The inverted domain appeared only at $U \leq -U_{cr}$ (path 45). A sufficiently 'big" domain acts as new matrix for the inverted one, appearing just below the tip at $U \leq -U_{cr}$ (path 45 and 56). The inverted domain size increases with further voltage decrease (path 56). At the point 6, the domain walls annihilate and the system returns to the initial state (path 61).

Note that the vertical asymmetry of the loop follows from the fact that the response of the nested domains (path 56) differs from the single one (path 1-3). Domain walls annihilate in point 6, then response coincides with the one from the initial state $d_{33}^{eff} = -72$ pm/V (path 4-0). The loop vertical asymmetry decreases under the maximal voltage increase, namely the loop 1-2-3-4-5-6 that corresponds to the maximal voltage of 10 V is strongly asymmetrical, whereas the loop 1-2-3-4’-5’-6 that corresponds to the maximal voltage of $10^3$ V becomes almost symmetrical.

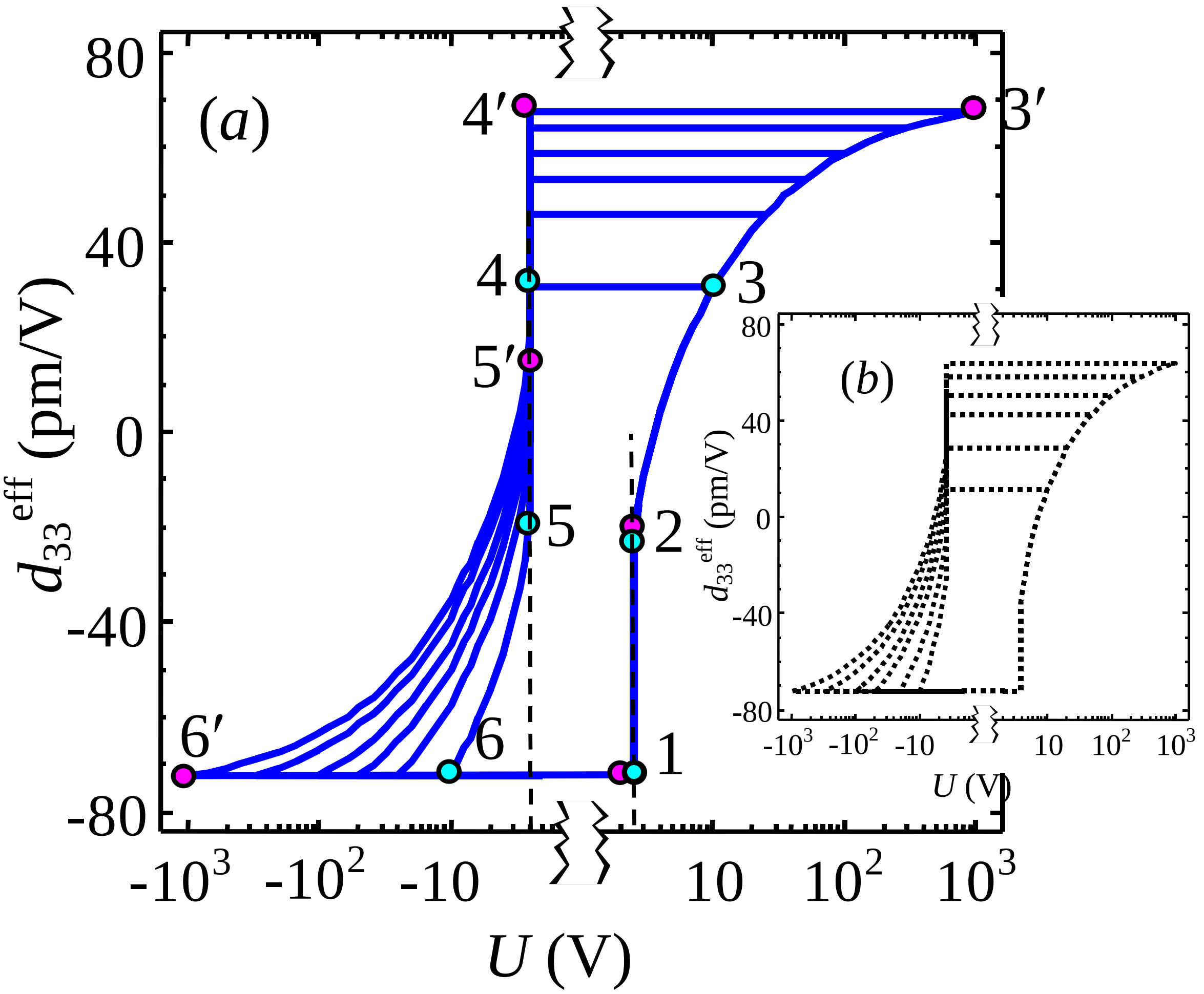

FIG. 14 (Color online) Piezoelectric response as the function of applied voltage for PZT6B at different maximal voltages 10, 25, 40, 100, 200 and $10^3$ V. Solid curves (a) represent EPCM approximation of the tip; dotted ones (b) correspond to the exact series for sphere-tip interaction energy. Material parameters and tip-surface characteristics are given in Fig.3; $d_{33} = 74.94$, $d_{31} = -28.67$ and $d_{15} = 135.59$ pm/V, $\sigma_S = -P_S$.

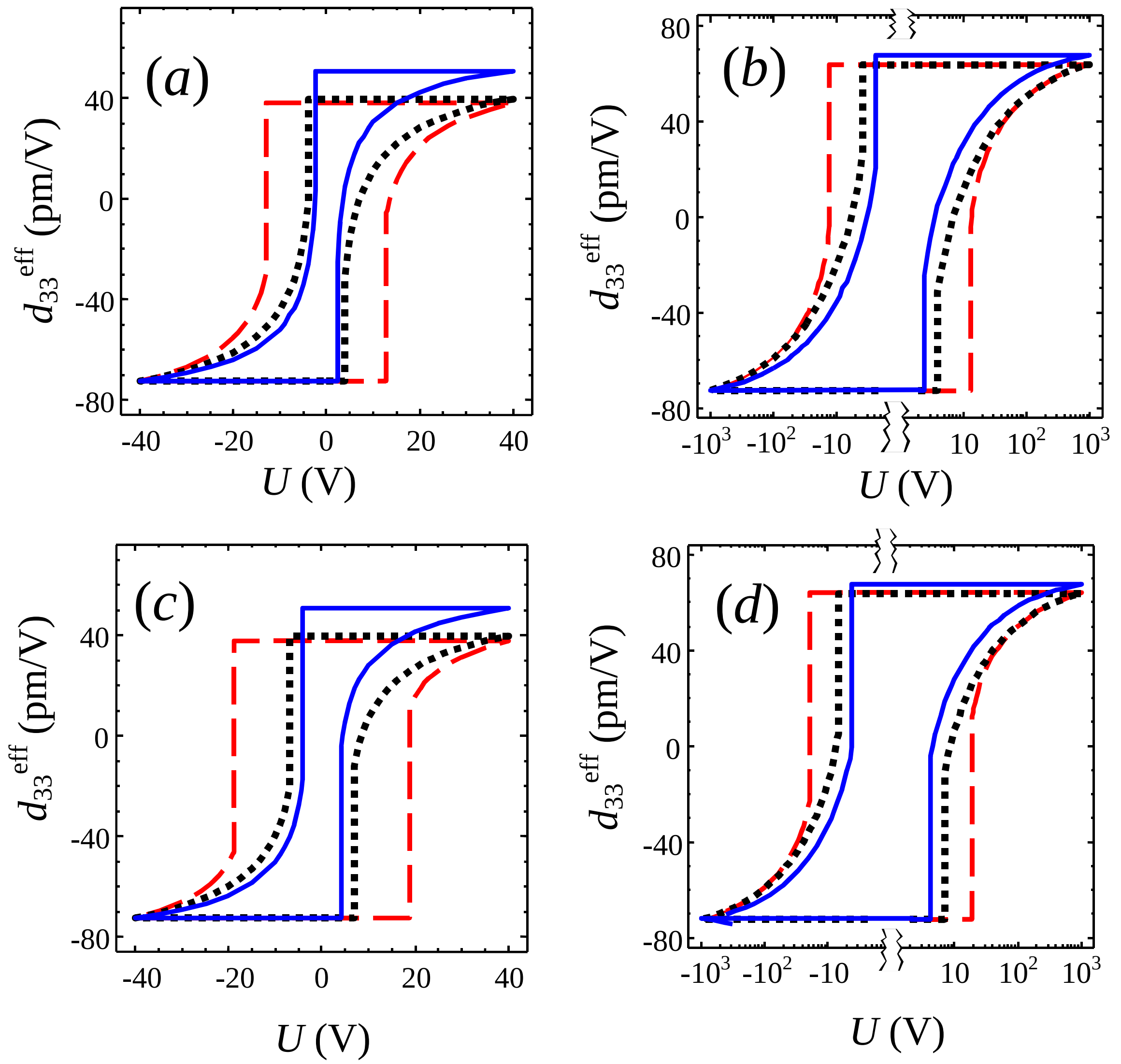

FIG. 15 (Color online) Piezoelectric response as the function of applied voltage for PZT6B and $\sigma_S = -P_S$ (a, b) and $\sigma_S = -0.5P_S$ (c, d) in linear (a,c) and logarithmic (b,d) scales. Solid curves represent EPCM approximation of the tip; dotted ones correspond to the exact series for sphere-tip interaction energy; dashed curves are the capacitance approximation. Material parameters and tip-surface characteristics are given in Fig.3; $d_{33} = 74.94$, $d_{31} = -28.66$ and $d_{15} = 135.59$ pm/V; whereas saturated value $d_{33}^{eff} = 72.5\,\text{pm/V}$ is depicted by arrows in parts (b) and (d).

Numerically, the results obtained within the EPCM of the tip at $R_d \to \infty$ can be well approximated by $d_{33}^{\mathrm{eff}} = d_\infty\left(1-\sqrt{U_0/U}\right)$. The difference with the one $d_{33}^{\mathrm{eff}} = d_\infty\left(1-U_0/U\right)$ obtained within the framework of 1D model[59] could be related to the dimensionality of the problem.

It is clear from Fig. 15 (b,d) that the modified point charge model gives the narrower loop that saturates more quickly than the exact series for sphere-tip interaction energy and moreover quicker than the CA model. This can be explained taking into account that the distance *d* between the effective point charge *Q* and the sample surface is smaller in $\kappa/\varepsilon_e \approx 6$ times than the first ones from the image charges caused by the tip with curvature $R_0$.

The influence of Debye screening radius $R_d$ on piezoelectric response is shown in Fig. 16 for PZT6B.

Despite the decrease of domain sizes the piezoresponse saturates much more quickly at small $R_d$ values (about 30 V for $R_d = 5\,\mathrm{nm}$) than at big ones (about 1 kV for $R_d = 500\ \mathrm{nm}$). The reason of this effect is explained by the quick vanishing of the tip potential at small $R_d$ radiuses (see Section IV.2.)

To summarize, the effect of surface screening and bulk Debye screening on piezoresponse loop shape, coercive voltage and saturation rate is the following:

(i) The surface screening strongly influences the domain nucleation and initial stage of growth. The coercive voltage (loop width) and nucleation voltages are controlled by $\sigma_S$ value. At the same time, piezoresponse weakly depends on $\sigma_S$ at high voltages, i.e. surface screening does not affect the saturation law (compare Fig. 15 and 16).

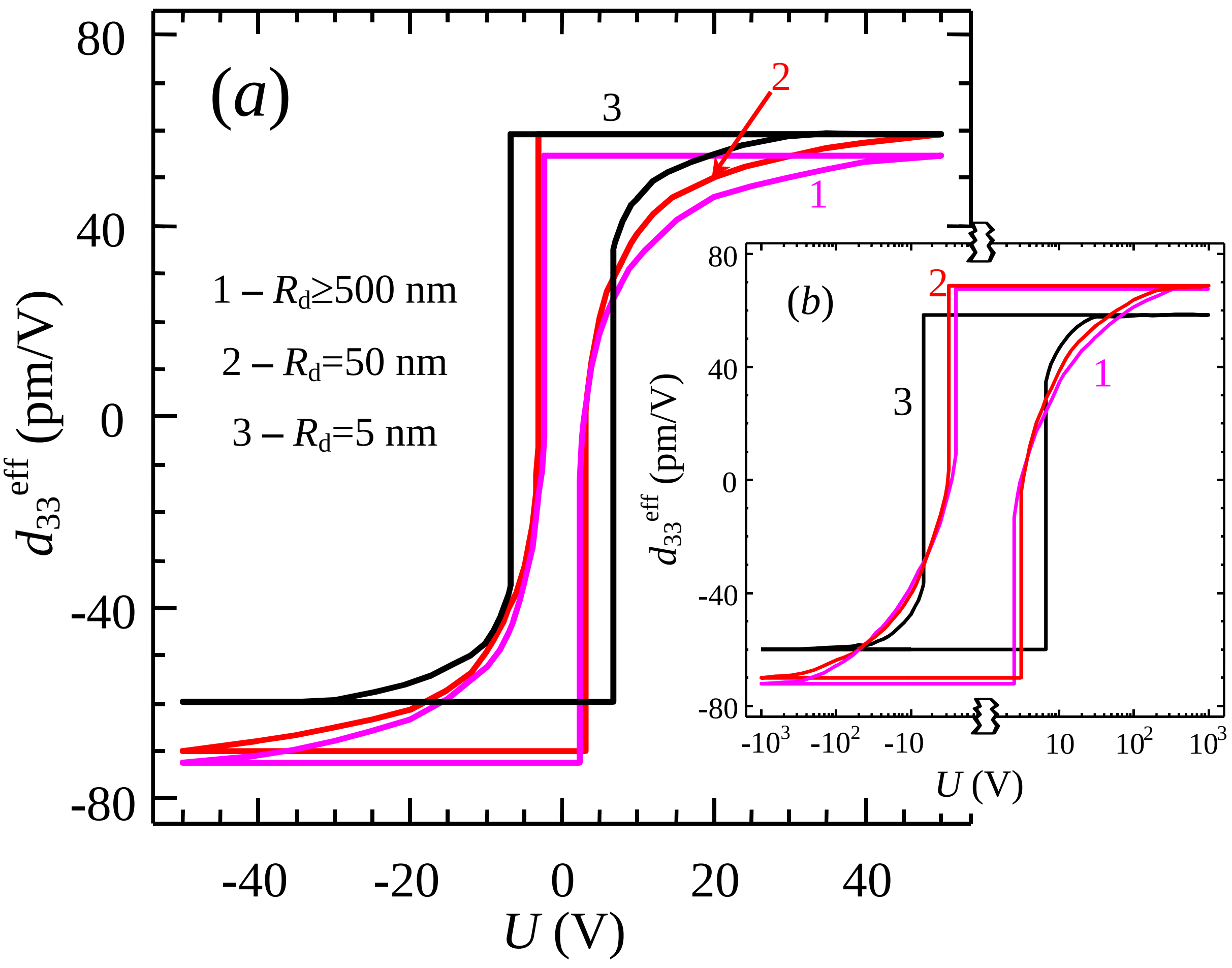

FIG. 16. (Color online) Piezoelectric response as the function of applied voltage in linear (a) and logarithmic (b) scales at different $R_d$ radius: ≥500 nm (curve 1), 50 nm (curve 2), 5 nm (curve 3). EPCM model with $R_0 = 50\,\text{nm}$, material parameters $\varepsilon_e = 81$ and $\kappa = 500$ correspond to PZT6B.

(ii) The Debye screening radius $R_d$ strongly influences the piezoresponse at high voltages and thus determines the saturation law (i.e. high voltage tails of hysteresis loop), whereas nucleation voltage depends on $R_d$ relatively weakly (compare Figs. 12(a) and (b,c)).

Thus, the effect of surface and Debye screening on piezoresponse loop shape is complementary with respect to domain nucleation and loop saturation behavior.

## V.3. Comparison with experiment

### V.3.1. Phenomenological loop behavior

In discussion of the agreement between theoretical and experimental results, we focus on two aspects, namely (a) nucleation bias and (b) overall loop shape.

(a) Nucleation bias.

Experimentally measured values of nucleation bias are determined by the activation energy for nucleation that decreases rapidly with applied bias. For experimentally measured values of 5 – 10 V the size of critical nucleus and effective activation energies are about 1-0.5 nm and 0.8 - 0.5eV respectively within the EPCM framework ( $\sigma_S = -P_S$ ). The corresponding nucleation times are $\tau = 2 - 2\cdot 10^{-5}$ s, making thermodynamic nucleation feasible even on ideal surface in the absence of defects.

(b) Loop shape

The most remarkable feature of the theoretical hysteresis loops in the weak pinning regime is that they are predicted to be extremely narrow and saturate rather slowly in ferroelectrics with large Debye lengths ( $R_d/r >> 1$ ). This behavior follows from the $1/r$ dependence of Green's function in 3D case, implying that the PFM signal will saturate to 90% of its final value when the domain diameter achieves 10 times the characteristic tip size (i.e. charge surface separation in the point charge model, or tip radius in the sphere plane one).

This behavior can be further understood given that in ferroelectric materials with $\gamma < 1$ the field is concentrated primarily in the surface region. At the same time, domains usually adopt prolate geometry ( $l >> r$ ). Hence, only the part of the domain close to the surface contributes to the PFM signal, and domain radius is the dominant length scale determining the PFS response. Relatively weak dependence of domain radius *r* on voltage *U* explains the slow saturation of the response since $R_d / r >> 1$. Note that in an elegant study by Kholkin *et al.*[28] domains imaged at different stages of the hysteresis loop illustrate that saturation is achieved only for domains of order of 200-300 nm, well above the tip size that can be estimated from the spatial resolution as ~20-30 nm.

Much faster saturation of the piezoresponse appears in ferroelectrics-semiconductors with small Debye radius due to faster decay of electric field. However, in this case, the current flow from tip to surface can significantly affect the PFM imaging.

### V.3.2. Implications for switching mechanism

Experimentally obtained hysteresis loops nearly always demonstrate much faster saturation then the loops predicted from thermodynamic theory. This behavior can be ascribed to several possible mechanisms, including (a) delayed domain nucleation (compared to thermodynamic model) due to poor tip-surface contact that leads to rapid jump from initial to final state, (b) finite conductivity and faster decay of electrostatic fields in the material, (c) kinetic effects on domain wall motion, and (d) surface screening and charge injection effects.

(a) Delayed nucleation

The activation barrier for nucleation is extremely sensitive to maximal electric field in the tip-surface junction region, which can be significantly reduced by surface adsorbates, quantum effects due to finite Thomas-Fermi length in tip material, polarization suppression at surfaces, etc. These factors are significantly less important for determining the fields at larger separation from contact, and hence affect primarily domain nucleation, rather then subsequent domain wall motion. This effect will result in sudden onset of switching, increasing the nucleation bias and rendering the loop squarer (see Fig. 17). However, the theory in Section IV suggest that to account for experimental observations, the nucleated domain size should be significantly larger than tip size, and nucleation should occur only for very high voltages. Given the generally good agreement between experimental and theoretical nucleation biases, we believe this effect does not explain experimental findings.

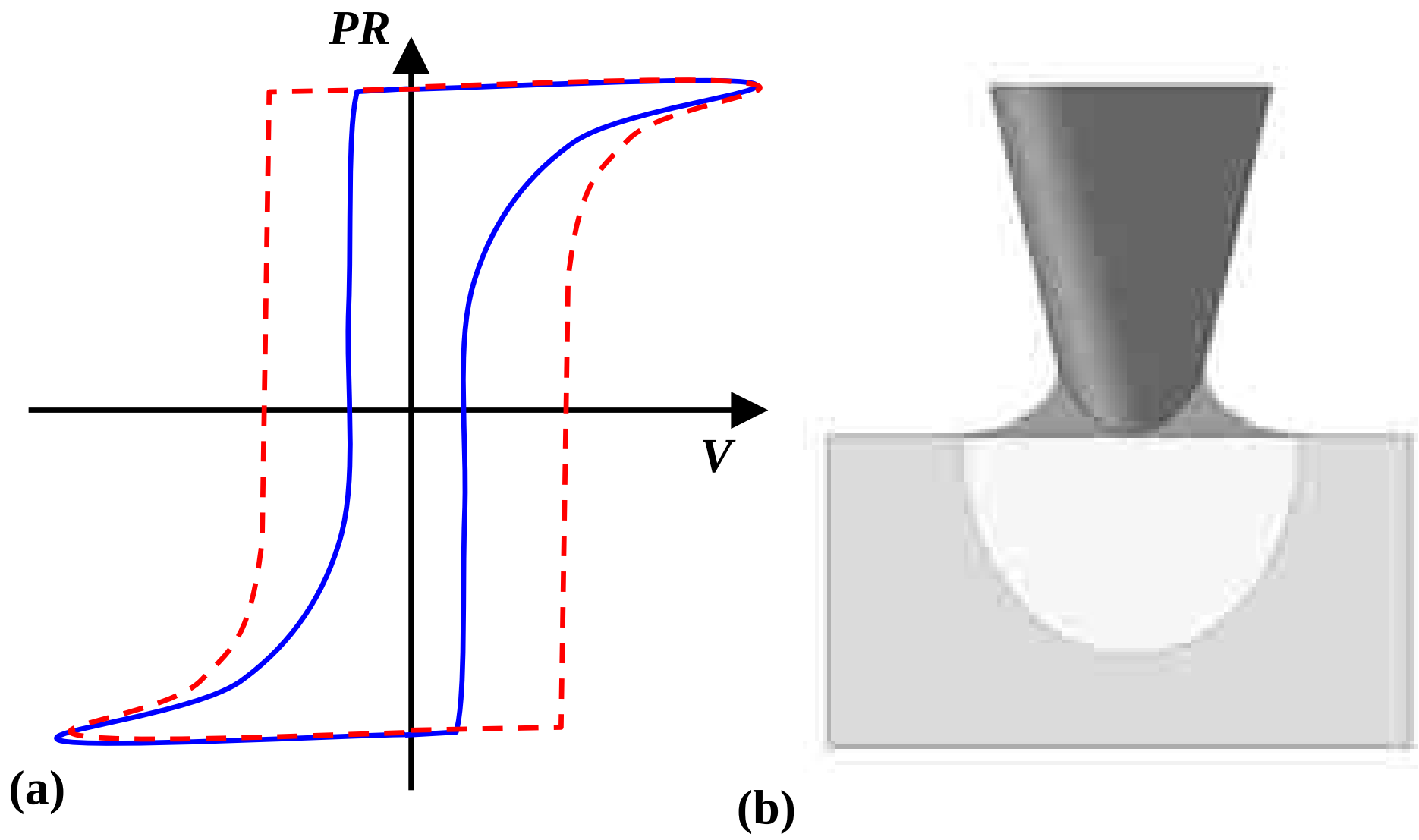

FIG. 17. (Color online) (a) Delayed nucleation will result in the broadening of the low-bias part of the hysteresis loop, effectively resulting in faster saturation. (b) The presence of the conductive water meniscus at the tip surface junction effectively broadens the radius of electrical tip-surface contact, while mechanical contact remains unaffected.

(b) Conductivity and finite Debye length

The second possible explanation for the observed behavior is finite conductivity of the sample and/or surrounding medium. In this case, screening by free carriers will result in cross-over from power law to exponential decay of electrostatic fields on the depth comparable to the Debye length (see Section V). This was shown to result in self-limiting effect in domain growth.[25,69] Given that in most materials studied to date Debye lengths are of the order of microns, this explanation cannot universally account for experimental observation.

(c) Domain wall motion kinetics

In realistic material, domain growth will be affected by the kinetics of domain wall motion. In the weak pinning regime, the domain size is close to the thermodynamically predicted, while in kinetic (strong pinning) regime the domain is significantly smaller. Both domain length and radius will grow slower then predicted by thermodynamic model. The detailed effect of pinning on domain shape is difficult to predict, since the field decays faster in z-direction, but at the same time surface pinning can dominate. In either case, pinning is likely to broaden hysteresis loop compared to thermodynamic shape, and is unlikely to affect nucleation, contrary to experimental observations.

(d) Surface conductivity effect

One of the most common factors in AFM experiments in ambient conditions is the formation and diffusion of charged species.[62, 63, 66, 67, 68] This behavior has been broadly reported for non-contact electrostatic measurements, which are directly sensitive to surface charges. At the same time, the screening charge effect has until recently been ignored in PFM studies, since this technique is not directly sensitive to surface charges. Recent studies by Buhlmann *et al.*[65] has illustrated that charge dynamics can explain anomalous domain switching and formation of bubble domains.

Here we note that surface charging can result in rapid broadening of the domain in radial direction, i.e. electrical radius of tip-surface contact grows with time. Given that only the part of the surface in contact with the tip results in cantilever deflection (i.e. electrical radius is much larger then mechanical radius), this will result in rapid saturation of the hysteresis loop. Note that similar effects were observed in e.g. dip-pen nanolithography[83] and the kinetics of this process is very similar to experimentally observed logarithmic kinetics of tip-induced domain growth. Estimating carrier mobility at $D \sim 10^{-11}$ $m^2/s$, diffusion length in 10 s is 1 micron. At the same time, the surface charge diffusion is unlikely to affect nucleation stage, since the latter is controlled by the region of maximal electric field directly at tip-surface junction. Also, charge dynamics is unlikely to affect PFM imaging, since the characteristic frequencies are significantly larger and at 100 kHz the diffusion length is 10 nm (compare to role of screening charges on dc and ac transport measurements by SPM).

To summarize, we believe that experimental results and theoretical models can be reconciled only if the radius of electrical contact is significantly larger then the radius of mechanical contact. This behavior is due to the migration of surface charged species ubiquitous on oxides surfaces in ambient. These processes are also likely to affect, if not control, the kinetics of domain growth processes observed in voltage-pulse methods.

## VI. Future prospects of PFS

Here we discuss the potential applications of PFS for mapping non-conventional switching behavior, namely nucleation centers and unusual polarization states. Since the seminal paper by Landauer,[26] it is recognized that switching in ferroelectric materials is controlled by switching centers that decrease the local activation energy for nucleation. Despite the dominance of this theory, no information on real-space imaging and structural aspects of these centers is available due to lack of appropriate imaging techniques. Here, we discuss a possible approach for the detection of these centers based on the Switching Spectroscopy PFM (SSPFM) method.

The direct evidence for the presence of switching center can be obtained from the measurement of nucleation bias in PFM. Since the center lowers the activation barrier for nucleation in the uniform field, the same effect can be anticipated in local measurements. Hence, mapping of nucleation bias in SSPFM provides an activation energy map for ferroelectric switching. The second effect in SSPFM data can be anticipated if the tip is positioned in the vicinity of nucleation center. In this case, the interaction of nucleated domain below the tip with the stress/electric field of defect should result in characteristic instability (or secondary domain nucleation). This behavior is predicted to result in rapid jumps in local hysteresis loops, providing the explanation of observed hysteresis loop fine structure.

Finally, recent work by Naumov has predicted the possibility of toroidal ferroelectric polarization states. In this case, anticipated hysteresis behavior is illustrated in Fig. 18. On the application of tip bias, the material transforms from toroidal state with zero electromechanical response to ferroelectric state, resulting in characteristic pinched loop shape. Thus, PFM spectroscopy can provide a pathway for identification of novel polarization orderings in low-dimensional ferroelectrics.

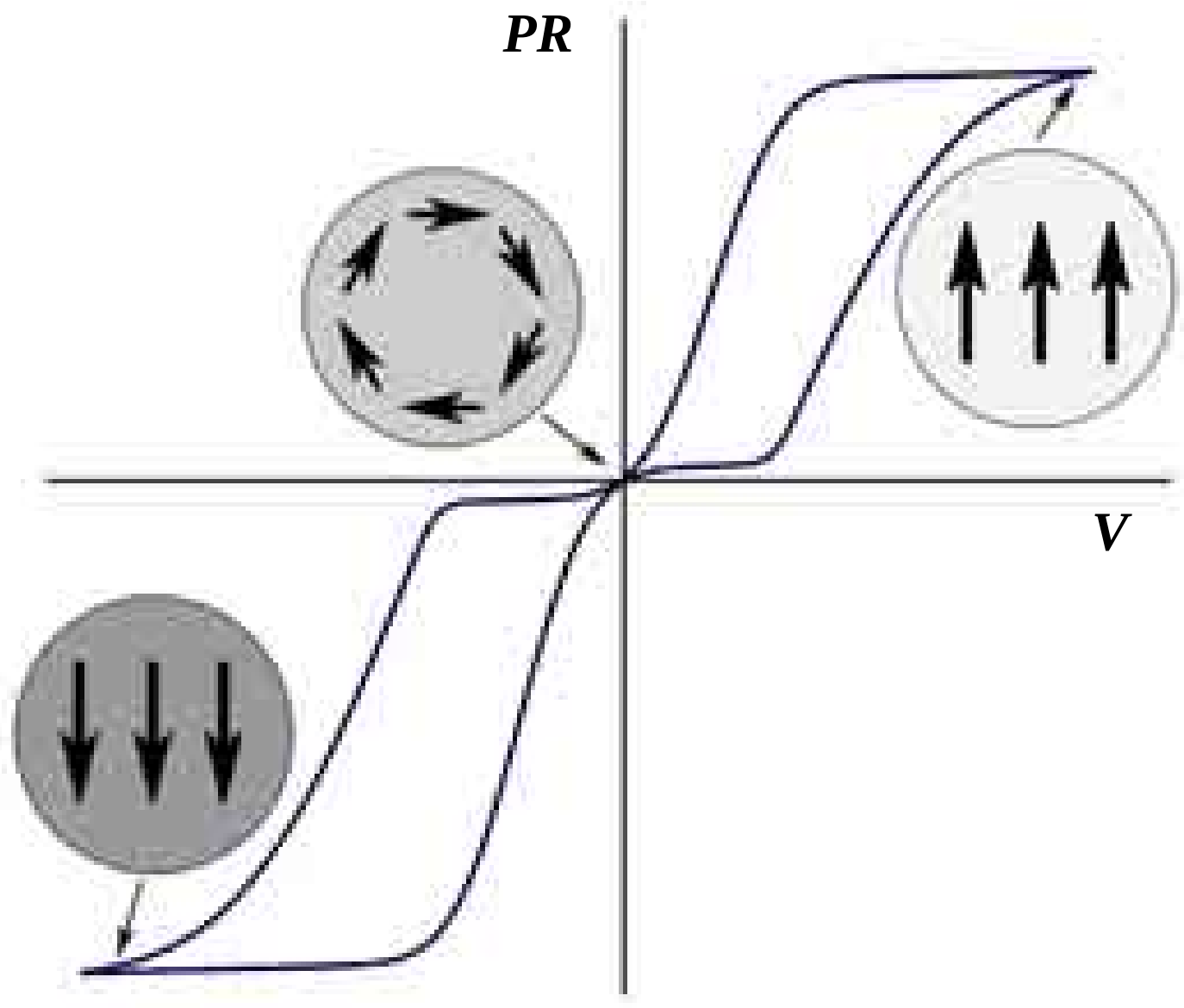

FIG. 18. (Color online) Anticipated evolution of polarization and electromechanical response for tip-induced switching in systems with toroidal polarization arrangements.

## VII. Conclusions

The hysteresis loop formation in PFM is analyzed in detail. The role of surface charges and finite Debye length on the thermodynamics of the switching process is elucidated. The general formalism relating parameters of domain to PFM signal is developed. This analysis is general and is applicable for modeling of arbitrary switching mechanisms, as well as for quantitative interpretation of PFS data. Actually we demonstrated that the effects of surface charges and Debye screening on piezoresponse loop are complementary with respect to domain nucleation and loop saturation behavior, namely:

(i) The surface charges strongly influence on the domain nucleation and initial stage of growth, whereas affect the high-voltage tail of hysteresis loop only weakly.

(ii) The value of Debye screening radius strongly influences on the piezoresponse behavior at high voltages and so determines the saturation law, whereas nucleation voltage is affected relatively weakly.

Comparison with experimental data indicates that experimental hysteresis loops saturate much faster than allowed by theory. The possible factors explaining this behavior, including domain wall pinning, finite conductivity, delayed nucleation, and surface charging are considered. Based on the comparison of experimental data and theoretical prediction, we believe that polarization switching processes are strongly mediated by the diffusion of surface charges generated in the tip-surface contact area. Surface charging increases the area of electrical contact, resulting in faster loop saturation, and also can account for experimentally observed logarithmic domain growth kinetics. Due to different time scales, the charges are unlikely to affect PFM imaging.

Finally, applicability of PFS for mapping nucleation centers and unusual polarization states in low-dimensional ferroelectrics are discussed and the characteristic identifying features in the loops are elucidated.

## Acknowledgements

Research supported by Oak Ridge National Laboratory, managed by UT-Battelle, LLC, for the U.S. Department of Energy under Contract DE-AC05-00OR22725.

## Appendix A. Pade approximations for the free energy.

a) The domain wall surface energy $\Phi_S(r,l)$ has the form:

$$\Phi_S(r,l)=\pi\psi_S\, l\, r\left(\frac{r}{l}+\frac{\arcsin\sqrt{1-r^2/l^2}}{\sqrt{1-r^2/l^2}}\right)\approx\frac{\pi^2\psi_S\, l\, d}{2}\left(1+\frac{2(r/l)^2}{4+\pi(r/l)}\right) \qquad \text{(A.1)}$$

Used in (A.1) Pade approximation $\left(x+\frac{\arcsin\sqrt{1-x^2}}{\sqrt{1-x^2}}\right)\approx\frac{\pi}{2}\left(1+\frac{2x^2}{4+\pi x}\right)$ fits exact expression

Eq. (A1) rather well, see Fig. 1A for details

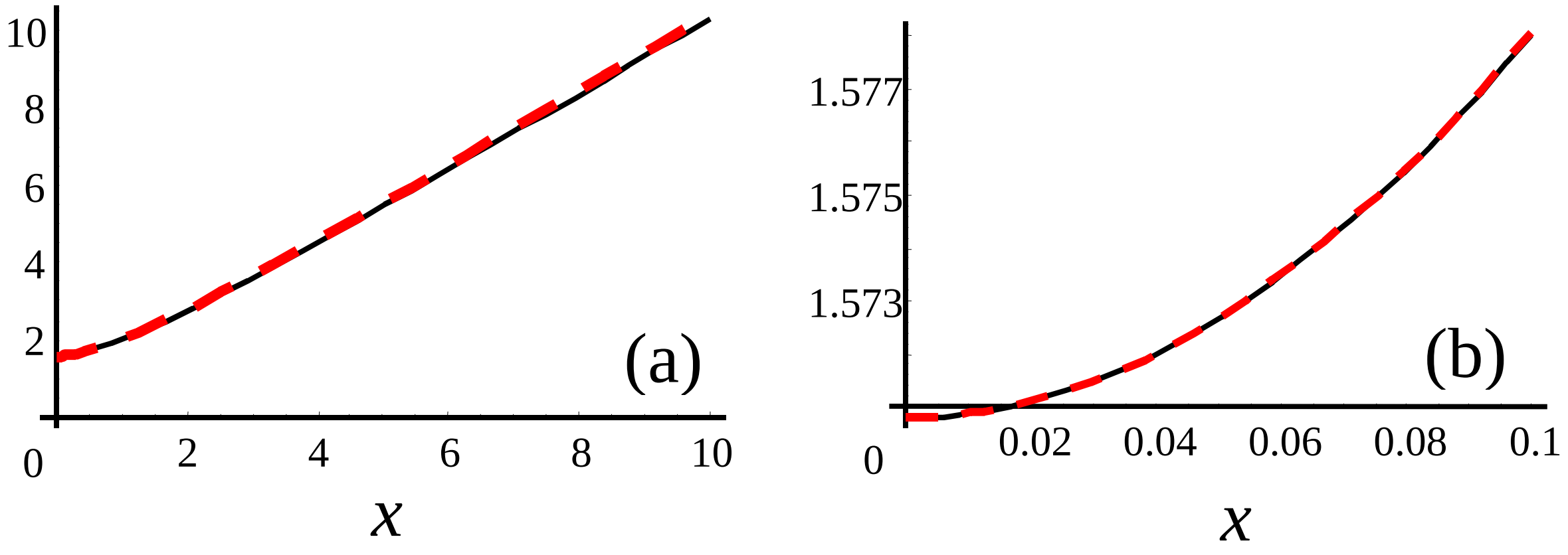

Fig. 1A (Color online) Comparison of exact expression $\left(x+\frac{\arcsin\sqrt{1-x^2}}{\sqrt{1-x^2}}\right)$ (solid curves) with proposed Pade approximation $\frac{\pi}{2}\left(1+\frac{2x^2}{4+\pi x}\right)$ (dashed curves).

b) Exact expression for interaction energy is

$$\Phi_U(r,l)\approx 4\pi\varepsilon_e U R_0\sum_{m=0}^{\infty} q_m\frac{R_d\left((\sigma_S-P_S)F_W(r,0,d-r_m)+2P_S F_W(r,l,d-r_m)\right)}{(\kappa+\varepsilon_e)R_d+2\kappa\sqrt{(d-r_m)^2+r^2}}$$

$$q_0=1,\quad q_m=\left(\frac{\kappa-\varepsilon_e}{\kappa+\varepsilon_e}\right)^m\frac{sh(\theta)}{sh((m+1)\theta)}, \qquad \text{(A.2)}$$

$$r_0=0,\quad r_m=R_0\frac{sh(m\theta)}{sh((m+1)\theta)},\quad ch(\theta)=\frac{d}{R_0},\qquad d=R_0+\Delta R$$

Exact expression for $F_W(d,l)$ is

$$F_W(r,l,d)=\begin{cases}\left(\dfrac{\sqrt{d^2+r^2}-(d+l/\gamma)}{(l/r\gamma)^2-1}+\dfrac{d}{\left((l/r\gamma)^2-1\right)\sqrt{1-(r\gamma/l)^2}}\times\right. \\ \left.\times\ln\left(\dfrac{d+(l/\gamma)\left(1-(r\gamma/l)^2\right)+(d+l/\gamma)\sqrt{1-(r\gamma/l)^2}}{d+\sqrt{1-(r\gamma/l)^2}\sqrt{r^2+d^2}}\right)\right) & at\ (\gamma r/l)<1 \\ \left(\dfrac{\sqrt{d^2+r^2}-(d+l/\gamma)}{(l/r\gamma)^2-1}+\dfrac{d}{\left((l/r\gamma)^2-1\right)\sqrt{(r\gamma/l)^2-1}}\times\right. \\ \left.\times\left(arctg\left(\dfrac{(d+l/\gamma)\sqrt{(r\gamma/l)^2-1}}{d-(l/\gamma)\left((r\gamma/l)^2-1\right)}\right)-arctg\left(\dfrac{\sqrt{d^2+r^2}\sqrt{(r\gamma/l)^2-1}}{d}\right)\right)\right) & at\ (\gamma r/l)\gtrsim 1\end{cases}$$

(A.3a)

Pade approximation $\dfrac{F_W(r,l,d)}{d}\approx\dfrac{r^2/d}{\sqrt{r^2+d^2}+d+(l/\gamma)}=\dfrac{-xy^2}{y+\left(\sqrt{1+y^2}+1\right)x}$, (A.3b)

where $x=(\gamma r/l)$ and $y=r/d$ provides a good approximation of exact Eq. (A.3a).

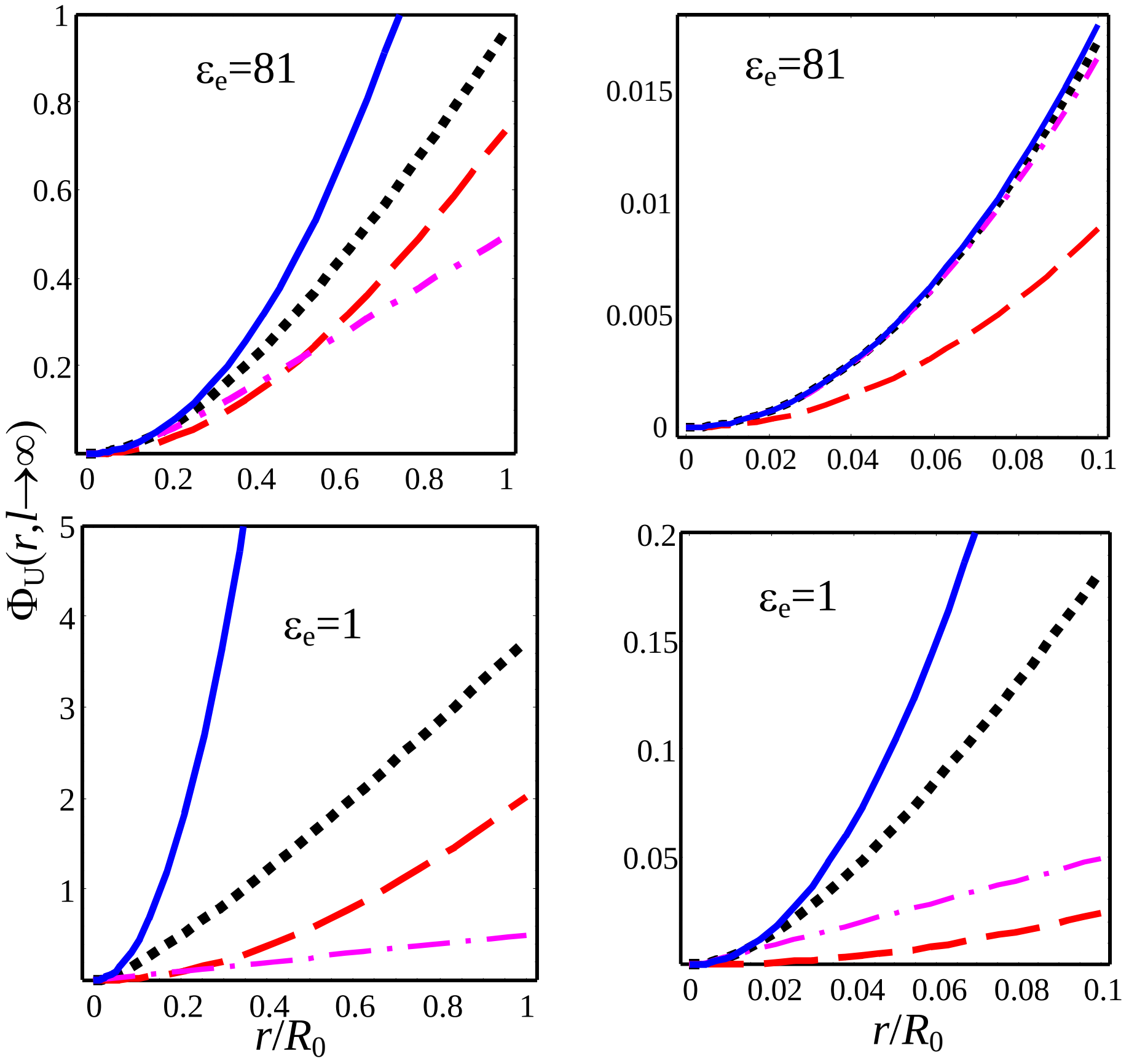

FIG. 2A. (Color online) Normalized interaction energy for PZT6B $\varepsilon_a = \varepsilon_c = 500$, tip in contact with sample $\Delta R = 0$. Dotted curves represent exact sphere – plane model (A.2), dashed ones – capacitance approximation, solid ones calculated for the effective charge model with effective distance $d = \varepsilon_e R_0/\kappa$, and dash-dotted ones correspond to the fitting $F_W(r,0,z) \approx r^2/2z$.

Finally, calculations in the effective point charge model for material with finite Debye length require the effective charge-surface distance $d$ for arbitrary $R_d$ values. The external electric field potential $V_S(\mathbf{r})$ created by the point charge $q$ localized in air (dielectric permittivity $\varepsilon_e$) in the point $r_0 = (0,0,-d)$, inside the semi-space $0 \le z \le \infty$ filled by isotropic semiconductor (dielectric permittivity $\varepsilon_i$ screening radius $R_d$ ) could be found from the boundary problem:

$$\begin{gathered}\Delta V_0(\mathbf{r})-V_0(\mathbf{r})=-\frac{Q}{\varepsilon_0\varepsilon_e}\delta(x,y,z+d), \qquad z\le 0,\\ \Delta V_S(\mathbf{r})-\frac{V_S(\mathbf{r})}{R_d^2}=0, \qquad 0\le z\le\infty,\\ V_0(z=0)=V_S(z=0), \qquad V_S(z\to\infty)=0\\ \left.\left(\varepsilon_e\frac{\partial V_0}{\partial z}-\kappa\frac{\partial V_S}{\partial z}\right)\right|_{z=0}=0,\end{gathered} \tag{A.4}$$

Here $V_S(\mathbf{r})$ is potential distribution inside the half-space ($z$ > 0) and $V_0(\mathbf{r})$ is potential distribution in air. The solution of (A.4) can be found with the help of Hankel integral transformation. Finally we obtained that:

$$V_0(\mathbf{r})=\frac{Q}{4\pi\varepsilon_0\varepsilon_e}\int_0^\infty dkJ_0\left(k\sqrt{x^2+y^2}\right)\begin{pmatrix}\exp(-k\cdot|z+d|)+\exp(k\cdot(z-d))\times\\ \times\dfrac{\varepsilon_e k-\kappa\sqrt{k^2+R_d^{-2}}}{\varepsilon_e k+\kappa\sqrt{k^2+R_d^{-2}}}\end{pmatrix} \tag{A.5}$$

$$V_S(\mathbf{r})=\frac{Q}{2\pi\varepsilon_0}\int_0^\infty dkJ_0\left(k\sqrt{x^2+y^2}\right)\exp(-dk)\frac{k\exp\left(-z\sqrt{k^2+R_d^{-2}}\right)}{\varepsilon_e k+\kappa\sqrt{k^2+R_d^{-2}}} \tag{A.6}$$

Then the effective distance $d$ should be found from the conditions that isopotential surface has the curvature $R_0$ in the point $(0,0,-h)$ of tip touching and $V_0(0,0,-h)=U$ (fixed tip potential). Namely, the effective distance $d$ should be found from the transcendental integral equation $\frac{1}{R_0}=-\frac{V_{0,xx}(0,0,-h)}{V_{0,z}(0,0,-h)}$. In the case $h\to 0$ we obtained the equation

$$F(d)=\frac{1}{R_0},\quad F(d)=\frac{\int_0^\infty \frac{dk\cdot\varepsilon_e k^3}{\varepsilon_e k+\kappa\sqrt{k^2+R_d^{-2}}}\exp(-d\,k)}{\int_0^\infty \frac{dk\cdot 2\kappa k\sqrt{k^2+R_d^{-2}}}{\varepsilon_e k+\kappa\sqrt{k^2+R_d^{-2}}}\exp(-d\,k)}. \tag{A.7}$$

In particular case $R_d\to\infty$ the effective distance $d=\varepsilon_e R_0/\kappa$. In all other cases the effective distance *d* should be found numerically from Eq.(A.7).

c) The exact expression for depolarization factor $n_D(r,l)$ is well known, namely

$$n_D(r,l)=\begin{cases}\frac{(r\gamma/l)^2}{1-(r\gamma/l)^2}\left(\frac{\operatorname{arcth}\left(\sqrt{1-(r\gamma/l)^2}\right)}{\sqrt{1-(r\gamma/l)^2}}-1\right), & at\quad \frac{l}{\gamma}\gtrsim r\\ \frac{(r\gamma/l)^2}{(r\gamma/l)^2-1}\left(1-\frac{\operatorname{arctg}\left(\sqrt{(r\gamma/l)^2-1}\right)}{\sqrt{(r\gamma/l)^2-1}}\right), & at\quad \frac{l}{\gamma}<r\end{cases} \tag{A.8}$$

The proposed approximation

$$n_D(x)\approx\frac{\sqrt{c_0^2+x^2}-c_0}{x+\pi/2}, \tag{A.9}$$

where $x=(r\gamma/l)$ is good even for derivatives at $x>0.05$.

The excess of the energy $\Delta\Phi_{DS}=\int_{\Sigma(z>0)}ds(\mathbf{P}_S\cdot\mathbf{n})\varphi_{DS}$ has been calculated using the surface charges potential

$$\varphi_{DS}(x,y,z)\approx\frac{(\sigma_S-P_S)r}{\varepsilon_0}\int_0^\infty dk\frac{J_0\left(k\sqrt{x^2+y^2}\right)J_1(kr)}{\varepsilon_e k+\kappa\sqrt{k^2+R_d^{-2}}}\exp\left(-z\sqrt{k^2+R_d^{-2}}\right). \tag{A.10}$$

Using Zommerfeld formulae

$$\frac{\exp\left(-\sqrt{x^2+y^2+z^2}/R_d\right)}{\sqrt{x^2+y^2+z^2}}=\int_0^\infty dk\frac{k\,J_0\left(k\sqrt{x^2+y^2}\right)}{\sqrt{k^2+R_d^{-2}}}\exp\left(-\sqrt{k^2+R_d^{-2}}\cdot|z|\right), \tag{A.11}$$

We obtain the estimate

$$\varphi_{DS} \sim \frac{(\sigma_S - P_S)r^2}{\varepsilon_0(\kappa + \varepsilon_e)} \frac{\exp\left(-\sqrt{x^2 + y^2 + z^2}/R_d\right)}{\sqrt{x^2 + y^2 + z^2}} \qquad \text{(A.12)}$$

valid at $r \neq 0$. Under these conditions we derived $\Delta\Phi_{DS}(r,l) \sim \frac{2P_S(\sigma_S - P_S)r^3}{1+(l/r\gamma)}$. Under the condition $l >> r$ this correction can be neglected.

## Appendix B. Approximate analytical expressions for domain sizes and nucleation threshold

The equilibrium sizes $\{r_{eq}, l_{eq}\}$ can be determined from the solution of equations

$$\begin{cases} \dfrac{\partial\Phi(r_{eq}, l_{eq})}{\partial r} = 0, \\ \dfrac{\partial\Phi(r_{eq}, l_{eq})}{\partial l} = 0. \end{cases} \tag{B.1}$$

Under the additional conditions $\partial^2\Phi(r,l)/\partial r^2 > 0$ and $\partial^2\Phi(r,l)/\partial l^2 > 0$ one obtains the absolute minimum $\{r_{eq}, l_{eq}\}$ from Eq. (9). Under the conditions $\partial^2\Phi(r,l)/\partial r^2 < 0$ or $\partial^2\Phi(r,l)/\partial l^2 < 0$ one obtains the saddle point $\{r_S, l_S\}$. In general case minimum and saddle points should be calculated numerically. Semi-quantitative analytical results could be obtained for prolate domains with $r < d$, $r << R_d$ and $(1 - \sigma_S/P_S) \approx 2$ from the free energy expansion as following:

$$\begin{gathered} \Phi(\tilde{r}, \tilde{l}) \approx f_U U \tilde{r}^2 \left( \left( \frac{\sigma_S}{P_S} - 1 \right) + \frac{2}{1 + \tilde{l}} \right) + f_D \left( \frac{\sigma_S}{P_S} - 1 \right)^2 \tilde{r}^3 + f_S \tilde{l} \tilde{r} \\ \tilde{r} = \frac{r}{d}, \qquad \tilde{l} = \frac{l}{2d\gamma}, \qquad \tilde{r} << 1, \qquad \tilde{r} << \tilde{l}, \qquad \frac{\sigma_S}{P_S} \approx -1, \\ f_U = \frac{R_d (C_t/\varepsilon_0) P_S d}{2(\kappa + \varepsilon_e) R_d + 4d\kappa}, \quad f_D = \frac{4P_S^2 d^3}{3\varepsilon_0 (\kappa + \varepsilon_e)}, \quad f_S = \pi^2 \psi_S d^2 \gamma \end{gathered} \tag{B.2}$$

Where in the modified point charge model $C_t \approx 4\pi\varepsilon_0\varepsilon_e R_0 \dfrac{\kappa + \varepsilon_e}{2\kappa}$ and $d = \varepsilon_e R_0/\kappa$. Using the dimensionless domain sizes $\tilde{r}$, $\tilde{l}$ and energies $f_{U,D,S}$ introduced in Eq. (B.2), we obtained from Eqs. (B.1)-(B.2) that:

$$\begin{cases} \tilde{l}(\tilde{r}) \approx \sqrt{\frac{2 f_U U}{f_S}}\sqrt{\tilde{r}} - 1 \\ 2\left(\frac{\sigma_S}{P_S} - 1\right) f_U U \tilde{r} + 3\sqrt{2 f_S f_U U}\sqrt{\tilde{r}} + 3 f_D \left(\frac{\sigma_S}{P_S} - 1\right)^2 \tilde{r}^2 - f_S = 0 \end{cases} \tag{B.3}$$

The latter fourth power equation for $\tilde{r}$ can be rewritten as the quadratic one for $U$ and solved parametrically along with the dependence $\tilde{l}(\tilde{r})$, namely

$$\begin{cases} U(\tilde{r}) \approx \left(\frac{3\sqrt{2 f_S} \pm \sqrt{18 f_S + 8(1-\sigma_S/P_S)\left(3 f_D (1-\sigma_S/P_S)^2 \tilde{r}^2 - f_S\right)}}{4\sqrt{f_U}(1-\sigma_S/P_S)\sqrt{\tilde{r}}}\right)^2 \\ \tilde{l}(\tilde{r}) \approx \sqrt{\frac{2 f_U U}{f_S}}\sqrt{\tilde{r}} - 1 \end{cases} \tag{B.4}$$

Note, that under the condition $(1-\sigma_S/P_S) > 0$, only the sign "+" before determinant in Eq.(B.4) corresponds to the physical states.

The energy barrier for domain nucleation in determined by the free energy value in the saddle point. The saddle point $\{U_S, \tilde{r}_S\}$ could be found from Eqs.(B.4) under the condition $\frac{\partial U(\tilde{r})}{\partial \tilde{r}} = 0$. When $\sigma_S = -P_S$, it was reduced to the cubic equation for $\tilde{r}_S^2$, namely $3 + \left(5 - 48\frac{f_D}{f_S}\tilde{r}_S^2\right)\sqrt{1 + 96\frac{f_D}{f_S}\tilde{r}_S^2} = 0$. Then $U_S(\tilde{r}_S)$ could be obtained from Eq. (B.4). We omit the general cumbersome expressions for $\tilde{r}_S^2$ and $U_S(\tilde{r}_S)$ for the sake of simplicity.

Using Eqs. (B.3) and condition $\Phi\left(\tilde{r}_{\min}, \tilde{l}_{\min}\right) = 0$ we derived the following approximate expressions for critical voltage $U_{cr}$ and sizes $\tilde{r}_{\min}$, $\tilde{l}_{\min}$

$$U_{cr} = \sqrt{\frac{f_D f_S}{1-\sigma_S/P_S}} \frac{1}{2 f_U} \frac{\left(3 + \sqrt{9 - 4(1-\sigma_S/P_S)}\right)^2}{\sqrt{3 - (1-\sigma_S/P_S) + \sqrt{9 - 4(1-\sigma_S/P_S)}}}, \tag{B.5}$$

$$\tilde{r}_{\min} = \sqrt{\frac{f_S}{f_D\left(1-\sigma_S/P_S\right)^3}}\sqrt{3-\left(1-\sigma_S/P_S\right)+\sqrt{9-4\left(1-\sigma_S/P_S\right)}}\ , \tag{B.6}$$

$$\tilde{l}_{\min} = \frac{3+\sqrt{9-4\left(1-\sigma_S/P_S\right)}}{1-\sigma_S/P_S}-1\,. \tag{B.7}$$

Using Eqs.(B.5)-(B.7) and the definitions of $\tilde{r}$ , $\tilde{l}$ , $f_{U,D,S}$ it is easy to obtain the functional dependencies of nucleation threshold on material parameters ant tip characteristics, namely:

$$U_{cr} = \frac{\left(\kappa+\varepsilon_e\right)R_d + 2d\kappa}{2\varepsilon_e R_d R_0 \left(\kappa+\varepsilon_e\right)/\kappa}\sqrt{\frac{\gamma\psi_S d^3}{\varepsilon_0\left(\kappa+\varepsilon_e\right)}}\frac{\left(3+\sqrt{9-4\left(1-\sigma_S/P_S\right)}\right)^2\Big/\sqrt{1-\sigma_S/P_S}}{\sqrt{3-\left(1-\sigma_S/P_S\right)+\sqrt{9-4\left(1-\sigma_S/P_S\right)}}}\ , \tag{B.8}$$

$$r_{\min} = \frac{\pi\sqrt{3\left(\kappa+\varepsilon_e\right)\varepsilon_0\gamma\psi_S}}{2P_S}\sqrt{d}\sqrt{\frac{\left(3-\left(1-\sigma_S/P_S\right)+\sqrt{9-4\left(1-\sigma_S/P_S\right)}\right)}{\left(1-\sigma_S/P_S\right)^3}}\ , \tag{B.9}$$

$$l_{\min} = \frac{2\gamma d}{1-\sigma_S/P_S}\left(2+\sigma_S/P_S+\sqrt{9-4\left(1-\sigma_S/P_S\right)}\right). \tag{B.10}$$

Note, that dependences (B.8)-(B.10) are derived from approximate free energy (B.2). Thus the approximate expressions give only semi-quantitative description of nucleation parameters.

## Appendix C. Intermediate states calculations

The Green's function components for isotropic semi-infinite half-plane is given by:[84], [85]

$$G_{ij}(x_1,x_2,x_3=0,\xi)=\begin{cases}\frac{1+\nu}{2\pi Y}\left(\frac{\delta_{ij}}{R}+\frac{(x_i-\xi_i)(x_j-\xi_j)}{R^3}+\right.\\ \left.+\frac{1-2\nu}{R+\xi_3}\left(\delta_{ij}-\frac{(x_i-\xi_i)(x_j-\xi_j)}{R(R+\xi_3)}\right)\right) & i,j\neq 3\\ \frac{(1+\nu)(x_i-\xi_i)}{2\pi Y}\left(\frac{-\xi_3}{R^3}\pm\frac{(1-2\nu)}{R(R+\xi_3)}\right) & \begin{matrix}\text{"+"} \sim j=1,2 \quad and \quad i=3\\ \text{"–"} \sim i=1,2 \quad and \quad j=3\end{matrix}\\ \frac{1+\nu}{2\pi Y}\left(\frac{2(1-\nu)}{R}+\frac{\xi_3^{\ 2}}{R^3}\right) & i=j=3\end{cases}$$

(C.1)

where $R=\sqrt{(x_1-\xi_1)^2+(x_2-\xi_2)^2+\xi_3^2}$ , $Y$ is Young's modulus and $\nu$ is Poisson ratio. The potential inside the transversally isotropic dielectric material produced by the point charge $Q$, at the distance $d$ above the surface, is

$$V_Q(\rho,z)=\frac{Q}{2\pi\varepsilon_0}\frac{1}{2\pi}\int_{-\infty}^{\infty}dk_x\int_{-\infty}^{\infty}dk_y\,\frac{\exp(-ik_x x-ik_y y)}{\varepsilon_e k+\kappa\sqrt{k^2+R_d^{-2}}}\exp\left(-\sqrt{k^2+R_d^{-2}}\cdot(z/\gamma)-k\,d\right)=$$
$$=\frac{Q}{2\pi\varepsilon_0}\int_0^{\infty}dk\,\frac{k\,J_0(k\rho)\exp\left(-\sqrt{k^2+R_d^{-2}}\cdot(z/\gamma)-k\,d\right)}{\varepsilon_e k+\kappa\sqrt{k^2+R_d^{-2}}} \qquad \text{(C.2)}$$

The displacement in the intermediate state is

$$u_3^i=V_Q(0,0)(d_{31}g_1+d_{15}g_2+d_{33}g_3), \qquad \text{(C.3a)}$$

$$g_i(\gamma,R_d,r,l)=f_i(\gamma,R_d)-2\,w_i(\gamma,R_d,r,l) \qquad \text{(C.3b)}$$

For the functions $f_i(\gamma,R_d)$ we obtained the series over the ratio $R_d/d$ and derived the following approximate expressions (18) from it, namely:

$$f_3(\gamma,R_d)=\begin{cases}-\dfrac{1+2\gamma}{(1+\gamma)^2}, & \dfrac{R_d}{d}>>1;\\ -1+\{4\}\gamma^2\dfrac{R_d^2}{d^2}, & \dfrac{R_d}{d}<<1.\end{cases}\approx-\frac{2\gamma\sqrt{1+\left(\dfrac{d}{2R_d}\right)^2}+1+\left(\dfrac{d}{2R_d}\right)^2}{\left(\sqrt{1+\left(\dfrac{d}{2R_d}\right)^2}+\gamma\right)^2}\tag{C.4a}$$

$$f_2(\gamma,R_d)=\begin{cases}-\dfrac{\gamma^2}{(1+\gamma)^2}, & \dfrac{R_d}{d}>>1,\\ -\{4\}\gamma^2\dfrac{R_d^2}{d^2}, & \dfrac{R_d}{d}<<1,\end{cases}\approx-\frac{\gamma^2}{\left(\sqrt{1+\left(\dfrac{d}{2R_d}\right)^2}+\gamma\right)^2}\tag{C.4b}$$

$$f_1(\gamma,R_d)=\begin{cases}-\dfrac{1+2(1+\gamma)\nu}{(1+\gamma)^2}, & \dfrac{R_d}{d}>>1,\\ \begin{pmatrix}-1-2\nu+\\ +4\gamma(1+\nu)\dfrac{R_d}{d}\end{pmatrix}, & \dfrac{R_d}{d}<<1\end{cases}\approx-\frac{2\gamma\nu\sqrt{1+\left(\dfrac{d}{2R_d}\right)^2}+(1+2\nu)\left(1+\left(\dfrac{d}{2R_d}\right)^2\right)}{\left(\sqrt{1+\left(\dfrac{d}{2R_d}\right)^2}+\gamma\right)^2}\tag{C.4c}$$

In particular case of cylindrical domain shape or prolate semi-ellipsoid ($r<<l$) the functions $w_i(\gamma,R_d,r)$ are given by the two-fold integrals, which allow Pade approximations:

$$w_3(\gamma,R_d,r)=-\frac{Q}{2\pi\varepsilon_0V_Q(0,0)}\int_0^\infty dq\int_0^\infty dk\,\frac{krJ_1(kr)J_0(qr)-qrJ_1(qr)J_0(kr)}{k^2-q^2}\frac{k\,q\exp(-k\,d)}{\varepsilon_e k+\kappa\sqrt{k^2+R_d^{-2}}}\times$$
$$\times\frac{\sqrt{k^2+R_d^{-2}}}{\gamma q+\sqrt{k^2+R_d^{-2}}}\frac{2\gamma q+\sqrt{k^2+R_d^{-2}}}{\gamma q+\sqrt{k^2+R_d^{-2}}}\approx\frac{f_3(\gamma,R_d)\sqrt{1+\left(\dfrac{r}{R_d}\right)^2}\left(\dfrac{r}{d}\right)}{\sqrt{1+\left(\dfrac{r}{R_d}\right)^2}\left(\dfrac{r}{d}\right)+\dfrac{B_3(\gamma)}{|f_3(\gamma,R_d)|}}\tag{C.5a}$$

$$w_2(\gamma,R_d,r)=-\frac{Q}{2\pi\varepsilon_0V_Q(0,0)}\int_0^\infty dq\int_0^\infty dk\,\frac{qr\,J_1(kr)J_0(qr)-kr\,J_1(qr)J_0(kr)}{k^2-q^2}\frac{k\,q\exp(-k\,d)}{\varepsilon_e k+\kappa\sqrt{k^2+R_d^{-2}}}\times$$
$$\times\frac{\gamma^2qk}{\left(\gamma q+\sqrt{k^2+R_d^{-2}}\right)^2}\approx\frac{f_2(\gamma,R_d)\sqrt{\left(1+\left(\dfrac{r}{R_d}\right)^2\right)^3}\left(\dfrac{r}{d}\right)}{\sqrt{\left(1+\left(\dfrac{r}{R_d}\right)^2\right)^3}\left(\dfrac{r}{d}\right)+\dfrac{B_2(\gamma)}{|f_2(\gamma,R_d)|}}\tag{C.5b}$$

$$w_1(\gamma, R_d, r) = -\frac{Q}{2\pi\varepsilon_0 V_Q(0,0)}\int_0^\infty dq\int_0^\infty dk \frac{krJ_1(kr)J_0(qr) - qrJ_1(qr)J_0(kr)}{k^2 - q^2}\frac{k\,q\exp(-k\,d)}{\varepsilon_e k + \kappa\sqrt{k^2 + R_d^{-2}}}\times$$

$$\times\frac{\sqrt{k^2 + R_d^{-2}}}{\gamma\,q + \sqrt{k^2 + R_d^{-2}}}\frac{2\nu\,\gamma\,q + (1 + 2\nu)\sqrt{k^2 + R_d^{-2}}}{\gamma\,q + \sqrt{k^2 + R_d^{-2}}} \approx (1 + 2\nu)w_3 - \frac{2(1+\nu)}{\gamma}\sqrt{1 + \left(\frac{d}{2R_d}\right)^2}w_2$$

(C.5c)

Constants $B_i(\gamma)$ depend solely on the dielectric anisotropy of material, namely:

$$B_3(\gamma) = \frac{3\pi}{32\gamma^2}\cdot {}_2F_1\left(\frac{3}{2},\frac{5}{2};4;1-\frac{1}{\gamma^2}\right),$$

$$B_2(\gamma) = \frac{3\pi}{32\gamma^2}\left[2\gamma^2\,{}_2F_1\left(\frac{1}{2},\frac{3}{2};3;1-\frac{1}{\gamma^2}\right) - {}_2F_1\left(\frac{3}{2},\frac{3}{2};4;1-\frac{1}{\gamma^2}\right)\right],$$ in particular $B_3(1) = B_2(1) = \frac{3\pi}{32}$.

Here ${}_2F_1(p,q;r;s)$ is the hypergeometric function.

---

[17]Y. Cho, S. Hashimoto, N. Odagawa, K. Tanaka and Y. Hiranaga, Nanotechnology **17,** S137–S141 (2006).

[18] M. Damjanovic, Hysteresis in Piezoelectric and Ferroelectric Materials, in The Science of Hysteresis Vol. 3, I. Mayergoyz and G. Bertolli (Eds), Elsevier 2005.

[19] Y. Ishibashi and H. orihara, Integr. Ferroelectr. **9**, 57 (1995).

[20] R. Waser, U. Bottger, and M. Grossman, in H. Ishiwara, M. Okuyama, and Y. Arimoto (Eds), Ferroelectric Random Access Memories, Topics Appl. Phys. **93**, 31 (2004).

[21] M. Abplanalp, Dr. Nat. Sci. thesis, Swiss Federal Institute of Technology, Zurich (2001).

[22] M. Molotskii, J.Appl. Phys. **93,** 6234 (2003).

[23] S.V. Kalinin, A. Gruverman B. J. Rodriguez, J. Shin, A. P. Baddorf, E. Karapetian and M. Kachanov, J. Appl. Phys. **97**, 074305 (2005).

[24] A.Yu. Emelyanov, Phys. Rev. B **71**, 132102 (2005).

[25] A.N. Morozovska and E.A. Eliseev, Phys. Rev. B. **73**, 104440 (2006).

[26] R. Landauer, J. Appl. Phys. **28**, 227 (1957).

[27] A. Gruverman, B.J. Rodriguez, C Dehoff, J.D. Waldrep, A.I. Kingon, R.J. Nemanich, and J.S. Cross, Appl. Phys. Lett. **87**, 082902 (2005).

[28] A.L. Kholkin, I.K. Bdikin, V.V. Shvartsman, A. Orlova, D. Kiselev, A.A. Bogomolov, and S.-H. Kim, *Mat. Res. Soc. Symp. Proc.* **839E**, O7.6 (2004).

[29] T. Tybell, P. Paruch, T. Giamarchi, and J.-M. Triscone, Phys. Rev. Lett. **89**, 097601 (2002).

[30] P. Paruch, T. Giamarchi, and J.-M. Triscone, Phys. Rev. Lett. **94**, 197601 (2005).

[31] M. Dawber, A. Gruverman and J. F. Scott, J. Phys.: Condens. Matter. **18**, L71 (2006).

[32] A. Agronin, Y. Rosenwaks, and G. Rosenman, Appl. Phys. Lett. **88**, 072911 (2006).

[46] W. Ma and D. Hesse, Appl. Phys. Lett. **84**, 2871 (2004).

[47] C. Harnagea, A. Pignolet, M. Alexe, D. Hesse, and U. Gosele, Appl. Phys. **A 70**, 261 (2000).

[48] D. Ricinschi, L. Mitoseriu, A. Stancu, P. Postolache, and M. Okuyama, Integrated Ferroelectrics **67**, 103 (2004).

[49] D. Ricinschi, M. Noda, M. Okuyama, Y. Ishibashi, M. Iwata, and L. Mitoseriu, J. Korean Phys. Soc. 42, S1232 (2003).

[50] D. Ricinschi and M. Okuyama, Integrated Ferroelectrics **50**, 159 (2002).

[51] A. Gruverman, B.J. Rodriguez, A.I. Kingon, R.J. Nemanich, J.S. Cross, and M. Tsukada, Appl. Phys. Lett. **82**, 3071 (2003).

[52] A. Gruverman, B.J. Rodriguez, A.I. Kingon, R.J. Nemanich, A.K. Tagantsev, J.S. Cross, and M. Tsukada, Appl. Phys. Lett. **83**, 728 (2003).

[53] A. Gruverman, B.J. Rodriguez, C Dehoff, J.D. Waldrep, A.I. Kingon, R.J. Nemanich, and J.S. Cross, Appl. Phys. Lett. **87**, 082902 (2005).

[54] C. Dehoff, B. J. Rodriguez, A. I. Kingon, R. J. Nemanich, A. Gruverman, and J. S. Cross, Rev. Sci. Instrum. **76**, 023708 (2005).

[55] M. Abplanalp, J. Fousek, and P. Gunter, Phys. Rev. Lett. **86**, 5799 (2001).

[56] C. Harnagea, Ph.D. Thesis, Halle 2001.

[57] B.J. Rodriguez *et al.*, to be submitted

[58] S. Jesse, H. N. Lee, and S. V. Kalinin, Rev. Sci. Instrum. **77**, 073702 (2006).

[59] S.V. Kalinin, A. Gruverman, D.A. Bonnell, Appl. Phys. Lett. **85**, 795 (2004).

[60] C.S. Ganpule, V. Nagarajan, H. Li, A.S. Ogale, D.E. Steinhauer, S. Aggarwal, E. Williams, R.Ramesh, P. DeWolf, Appl. Phys. Lett. **77**, 292 (2000)

[81] E.A. Eliseev, S.V. Kalinin, S. Jesse, S.L. Bravina, and A.N. Morozovska, cond-mat/0607543 (unpublished).

[82] S. Jesse, A.P. Baddorf, and S.V. Kalinin, Appl. Phys. Lett. **88**, 062908 (2006).

[83] P. E. Sheehan and L. J. Whitman, Phys. Rev. Lett. **88**, 156104 (2002).

[84] L.D. Landau and E.M. Lifshitz, *Theory of Elasticity*. Theoretical Physics, Vol. 7 (Butterworth-Heinemann, Oxford, U.K., 1998).

[85] A. I. Lur'e, *Three-dimensional problems of the theory of elasticity* (Interscience Publishers, New York, 1964).